\documentstyle[10pt,epsf,epsfig,hangcaption,xspace,amssymb,amsfonts,amsmath,amsthm,cite,
dp_delphititle,axodraw,lineno]{dp_delphi}
\makeindex
\def\DpPaperGroup{EP}
\def\DpPaperRef{2003-009}
\def\DpDate{6 February 2003}
\def\DpAuthors{DELPHI Collaboration}
\def\DpSubmit{(Accepted by Eur.Phys.J.C)}
\def\DpTitle{\boldmath $ZZ$ production in $e^+ e^-$ interactions
at $\sqrt{s} = 183 - 209$ GeV}
\def\DpComment{ }
\def\DpEMail{ }

\newcommand{\ba}{\begin{array}}
\newcommand{\ea}{\end{array}}
\newcommand{\bc}{\begin{center}}
\newcommand{\ec}{\end{center}}
\newcommand{\bt}{\begin{tabular}}
\newcommand{\et}{\end{tabular}}
\newcommand{\beq}{\begin{eqnarray}}
\newcommand{\eeq}{\end{eqnarray}}
\newcommand{\bes}{\begin{eqnarray*}}
\newcommand{\ees}{\end{eqnarray*}}
\begin{document}
\makeatletter
\newcount\@tempcntc
\def\@citex[#1]#2{\if@filesw\immediate\write\@auxout{\string\citation{#2}}\fi
\@tempcnta\z@\@tempcntb\m@ne\def\@citea{}\@cite{\@for\@citeb:=#2\do
{\@ifundefined
{b@\@citeb}{\@citeo\@tempcntb\m@ne\@citea\def\@citea{,}{\bf ?}\@warning
{Citation `\@citeb' on page \thepage \space undefined}}%
{\setbox\z@\hbox{\global\@tempcntc0\csname b@\@citeb\endcsname\relax}%
\ifnum\@tempcntc=\z@ \@citeo\@tempcntb\m@ne
\@citea\def\@citea{,}\hbox{\csname b@\@citeb\endcsname}%
\else
\advance\@tempcntb\@ne
\ifnum\@tempcntb=\@tempcntc
\else\advance\@tempcntb\m@ne\@citeo
\@tempcnta\@tempcntc\@tempcntb\@tempcntc\fi\fi}}\@citeo}{#1}}
\def\@citeo{\ifnum\@tempcnta>\@tempcntb\else\@citea\def\@citea{,}%
\ifnum\@tempcnta=\@tempcntb\the\@tempcnta\else
{\advance\@tempcnta\@ne\ifnum\@tempcnta=\@tempcntb \else \def\@citea{--}\fi
\advance\@tempcnta\m@ne\the\@tempcnta\@citea\the\@tempcntb}\fi\fi}
\makeatletter
\newcount\@tempcntc
\def\@citex[#1]#2{\if@filesw\immediate\write\@auxout{\string\citation{#2}}\fi
  \@tempcnta\z@\@tempcntb\m@ne\def\@citea{}\@cite{\@for\@citeb:=#2\do
    {\@ifundefined
       {b@\@citeb}{\@citeo\@tempcntb\m@ne\@citea\def\@citea{,}{\bf ?}\@warning
       {Citation `\@citeb' on page \thepage \space undefined}}%
    {\setbox\z@\hbox{\global\@tempcntc0\csname b@\@citeb\endcsname\relax}%
     \ifnum\@tempcntc=\z@ \@citeo\@tempcntb\m@ne
       \@citea\def\@citea{,}\hbox{\csname b@\@citeb\endcsname}%
     \else
      \advance\@tempcntb\@ne
      \ifnum\@tempcntb=\@tempcntc
      \else\advance\@tempcntb\m@ne\@citeo
      \@tempcnta\@tempcntc\@tempcntb\@tempcntc\fi\fi}}\@citeo}{#1}}
\def\@citeo{\ifnum\@tempcnta>\@tempcntb\else\@citea\def\@citea{,}%
  \ifnum\@tempcnta=\@tempcntb\the\@tempcnta\else
   {\advance\@tempcnta\@ne\ifnum\@tempcnta=\@tempcntb \else \def\@citea{--}\fi
    \advance\@tempcnta\m@ne\the\@tempcnta\@citea\the\@tempcntb}\fi\fi}
 
\makeatother
\begin{titlepage}
\pagenumbering{roman}
\CERNpreprint{\DpPaperGroup}{\DpPaperRef} 
\date{{\small\DpDate}} 
\title{\DpTitle} 
\address{\DpAuthors} 
\begin{shortabs} 
\noindent
\noindent

Measurements of on-shell $ZZ$ production are described, using 
data from the DELPHI experiment at LEP in $e^+ e^-$
collisions at centre-of-mass energies between 183 and 209 GeV,
corresponding to an integrated luminosity of about 665~pb$^{-1}$. 
Results obtained in each of the final states
$q \bar{q} q \bar{q}$,
$\nu \bar{\nu} q \bar{q}$, 
$\mu^+ \mu^- q \bar{q}$, 
$e^+e^- q \bar{q}$, 
$\tau^+ \tau^- q \bar{q}$, 
$l^+l^-l^+l^-$, and
$\nu \bar{\nu} l^+l^-$ (with $l=e,\mu$)
are presented. 
The measured production cross-sections
are consistent with the Standard Model expectations.
These results update and supersede those already published 
at 183 and 189 GeV. 
\end{shortabs}
\vfill
\begin{center}
\DpSubmit \ \\ 
\DpComment \ \\
\DpEMail \ \\
\end{center}
\vfill
\clearpage
\headsep 10.0pt
\addtolength{\textheight}{10mm}
\addtolength{\footskip}{-5mm}
\begingroup
%
\newcommand{\DpName}[2]{\hbox{#1$^{\ref{#2}}$},\hfill}
\newcommand{\DpNameTwo}[3]{\hbox{#1$^{\ref{#2},\ref{#3}}$},\hfill}
\newcommand{\DpNameThree}[4]{\hbox{#1$^{\ref{#2},\ref{#3},\ref{#4}}$},\hfill}
\newskip\Bigfill \Bigfill = 0pt plus 1000fill
\newcommand{\DpNameLast}[2]{\hbox{#1$^{\ref{#2}}$}\hspace{\Bigfill}}
%
\footnotesize
\noindent
\DpName{J.Abdallah}{LPNHE}
\DpName{P.Abreu}{LIP}
\DpName{W.Adam}{VIENNA}
\DpName{P.Adzic}{DEMOKRITOS}
\DpName{T.Albrecht}{KARLSRUHE}
\DpName{T.Alderweireld}{AIM}
\DpName{R.Alemany-Fernandez}{CERN}
\DpName{T.Allmendinger}{KARLSRUHE}
\DpName{P.P.Allport}{LIVERPOOL}
\DpName{U.Amaldi}{MILANO2}
% ...
\normalsize
\endgroup
\titlefoot{Department of Physics and Astronomy, Iowa State
     University, Ames IA 50011-3160, USA
    \label{AMES}}
\titlefoot{Physics Department, Universiteit Antwerpen,
     Universiteitsplein 1, B-2610 Antwerpen, Belgium \\
     \indent~~and IIHE, ULB-VUB,
     Pleinlaan 2, B-1050 Brussels, Belgium \\
     \indent~~and Facult\'e des Sciences,
     Univ. de l'Etat Mons, Av. Maistriau 19, B-7000 Mons, Belgium
    \label{AIM}}
\titlefoot{Physics Laboratory, University of Athens, Solonos Str.
     104, GR-10680 Athens, Greece
    \label{ATHENS}}
\titlefoot{Department of Physics, University of Bergen,
     All\'egaten 55, NO-5007 Bergen, Norway
    \label{BERGEN}}
\titlefoot{Dipartimento di Fisica, Universit\`a di Bologna and INFN,
     Via Irnerio 46, IT-40126 Bologna, Italy
    \label{BOLOGNA}}
\titlefoot{Centro Brasileiro de Pesquisas F\'{\i}sicas, rua Xavier Sigaud 150,
     BR-22290 Rio de Janeiro, Brazil \\
     \indent~~and Depto. de F\'{\i}sica, Pont. Univ. Cat\'olica,
     C.P. 38071 BR-22453 Rio de Janeiro, Brazil \\
     \indent~~and Inst. de F\'{\i}sica, Univ. Estadual do Rio de Janeiro,
     rua S\~{a}o Francisco Xavier 524, Rio de Janeiro, Brazil
    \label{BRASIL}}
\titlefoot{Coll\`ege de France, Lab. de Physique Corpusculaire, IN2P3-CNRS,
     FR-75231 Paris Cedex 05, France
    \label{CDF}}
\titlefoot{CERN, CH-1211 Geneva 23, Switzerland
    \label{CERN}}
\titlefoot{Institut de Recherches Subatomiques, IN2P3 - CNRS/ULP - BP20,
     FR-67037 Strasbourg Cedex, France
    \label{CRN}}
\titlefoot{Now at DESY-Zeuthen, Platanenallee 6, D-15735 Zeuthen, Germany
    \label{DESY}}
\titlefoot{Institute of Nuclear Physics, N.C.S.R. Demokritos,
     P.O. Box 60228, GR-15310 Athens, Greece
    \label{DEMOKRITOS}}
\titlefoot{FZU, Inst. of Phys. of the C.A.S. High Energy Physics Division,
     Na Slovance 2, CZ-180 40, Praha 8, Czech Republic
    \label{FZU}}
\titlefoot{Dipartimento di Fisica, Universit\`a di Genova and INFN,
     Via Dodecaneso 33, IT-16146 Genova, Italy
    \label{GENOVA}}
\titlefoot{Institut des Sciences Nucl\'eaires, IN2P3-CNRS, Universit\'e
     de Grenoble 1, FR-38026 Grenoble Cedex, France
    \label{GRENOBLE}}
\titlefoot{Helsinki Institute of Physics, P.O. Box 64,
     FIN-00014 University of Helsinki, Finland
    \label{HELSINKI}}
\titlefoot{Joint Institute for Nuclear Research, Dubna, Head Post
     Office, P.O. Box 79, RU-101 000 Moscow, Russian Federation
    \label{JINR}}
\titlefoot{Institut f\"ur Experimentelle Kernphysik,
     Universit\"at Karlsruhe, Postfach 6980, DE-76128 Karlsruhe,
     Germany
    \label{KARLSRUHE}}
\titlefoot{Institute of Nuclear Physics,Ul. Kawiory 26a,
     PL-30055 Krakow, Poland
    \label{KRAKOW1}}
\titlefoot{Faculty of Physics and Nuclear Techniques, University of Mining
     and Metallurgy, PL-30055 Krakow, Poland
    \label{KRAKOW2}}
\titlefoot{Universit\'e de Paris-Sud, Lab. de l'Acc\'el\'erateur
     Lin\'eaire, IN2P3-CNRS, B\^{a}t. 200, FR-91405 Orsay Cedex, France
    \label{LAL}}
\titlefoot{School of Physics and Chemistry, University of Lancaster,
     Lancaster LA1 4YB, UK
    \label{LANCASTER}}
\titlefoot{LIP, IST, FCUL - Av. Elias Garcia, 14-$1^{o}$,
     PT-1000 Lisboa Codex, Portugal
    \label{LIP}}
\titlefoot{Department of Physics, University of Liverpool, P.O.
     Box 147, Liverpool L69 3BX, UK
    \label{LIVERPOOL}}
\titlefoot{Dept. of Physics and Astronomy, Kelvin Building,
     University of Glasgow, Glasgow G12 8QQ
    \label{GLASGOW}}
\titlefoot{LPNHE, IN2P3-CNRS, Univ.~Paris VI et VII, Tour 33 (RdC),
     4 place Jussieu, FR-75252 Paris Cedex 05, France
    \label{LPNHE}}
\titlefoot{Department of Physics, University of Lund,
     S\"olvegatan 14, SE-223 63 Lund, Sweden
    \label{LUND}}
\titlefoot{Universit\'e Claude Bernard de Lyon, IPNL, IN2P3-CNRS,
     FR-69622 Villeurbanne Cedex, France
    \label{LYON}}
\titlefoot{Dipartimento di Fisica, Universit\`a di Milano and INFN-MILANO,
     Via Celoria 16, IT-20133 Milan, Italy
    \label{MILANO}}
\titlefoot{Dipartimento di Fisica, Univ. di Milano-Bicocca and
     INFN-MILANO, Piazza della Scienza 2, IT-20126 Milan, Italy
    \label{MILANO2}}
\titlefoot{IPNP of MFF, Charles Univ., Areal MFF,
     V Holesovickach 2, CZ-180 00, Praha 8, Czech Republic
    \label{NC}}
\titlefoot{NIKHEF, Postbus 41882, NL-1009 DB
     Amsterdam, The Netherlands
    \label{NIKHEF}}
\titlefoot{National Technical University, Physics Department,
     Zografou Campus, GR-15773 Athens, Greece
    \label{NTU-ATHENS}}
\titlefoot{Physics Department, University of Oslo, Blindern,
     NO-0316 Oslo, Norway
    \label{OSLO}}
\titlefoot{Dpto. Fisica, Univ. Oviedo, Avda. Calvo Sotelo
     s/n, ES-33007 Oviedo, Spain
    \label{OVIEDO}}
\titlefoot{Department of Physics, University of Oxford,
     Keble Road, Oxford OX1 3RH, UK
    \label{OXFORD}}
\titlefoot{Dipartimento di Fisica, Universit\`a di Padova and
     INFN, Via Marzolo 8, IT-35131 Padua, Italy
    \label{PADOVA}}
\titlefoot{Rutherford Appleton Laboratory, Chilton, Didcot
     OX11 OQX, UK
    \label{RAL}}
\titlefoot{Dipartimento di Fisica, Universit\`a di Roma II and
     INFN, Tor Vergata, IT-00173 Rome, Italy
    \label{ROMA2}}
\titlefoot{Dipartimento di Fisica, Universit\`a di Roma III and
     INFN, Via della Vasca Navale 84, IT-00146 Rome, Italy
    \label{ROMA3}}
\titlefoot{DAPNIA/Service de Physique des Particules,
     CEA-Saclay, FR-91191 Gif-sur-Yvette Cedex, France
    \label{SACLAY}}
\titlefoot{Instituto de Fisica de Cantabria (CSIC-UC), Avda.
     los Castros s/n, ES-39006 Santander, Spain
    \label{SANTANDER}}
\titlefoot{Inst. for High Energy Physics, Serpukov
     P.O. Box 35, Protvino, (Moscow Region), Russian Federation
    \label{SERPUKHOV}}
\titlefoot{J. Stefan Institute, Jamova 39, SI-1000 Ljubljana, Slovenia
     and Laboratory for Astroparticle Physics,\\
     \indent~~Nova Gorica Polytechnic, Kostanjeviska 16a, SI-5000 Nova Gorica, Slovenia, \\
     \indent~~and Department of Physics, University of Ljubljana,
     SI-1000 Ljubljana, Slovenia
    \label{SLOVENIJA}}
\titlefoot{Fysikum, Stockholm University,
     Box 6730, SE-113 85 Stockholm, Sweden
    \label{STOCKHOLM}}
\titlefoot{Dipartimento di Fisica Sperimentale, Universit\`a di
     Torino and INFN, Via P. Giuria 1, IT-10125 Turin, Italy
    \label{TORINO}}
\titlefoot{INFN,Sezione di Torino, and Dipartimento di Fisica Teorica,
     Universit\`a di Torino, Via P. Giuria 1,\\
     \indent~~IT-10125 Turin, Italy
    \label{TORINOTH}}
\titlefoot{Dipartimento di Fisica, Universit\`a di Trieste and
     INFN, Via A. Valerio 2, IT-34127 Trieste, Italy \\
     \indent~~and Istituto di Fisica, Universit\`a di Udine,
     IT-33100 Udine, Italy
    \label{TU}}
\titlefoot{Univ. Federal do Rio de Janeiro, C.P. 68528
     Cidade Univ., Ilha do Fund\~ao
     BR-21945-970 Rio de Janeiro, Brazil
    \label{UFRJ}}
\titlefoot{Department of Radiation Sciences, University of
     Uppsala, P.O. Box 535, SE-751 21 Uppsala, Sweden
    \label{UPPSALA}}
\titlefoot{IFIC, Valencia-CSIC, and D.F.A.M.N., U. de Valencia,
     Avda. Dr. Moliner 50, ES-46100 Burjassot (Valencia), Spain
    \label{VALENCIA}}
\titlefoot{Institut f\"ur Hochenergiephysik, \"Osterr. Akad.
     d. Wissensch., Nikolsdorfergasse 18, AT-1050 Vienna, Austria
    \label{VIENNA}}
\titlefoot{Inst. Nuclear Studies and University of Warsaw, Ul.
     Hoza 69, PL-00681 Warsaw, Poland
    \label{WARSZAWA}}
\titlefoot{Fachbereich Physik, University of Wuppertal, Postfach
     100 127, DE-42097 Wuppertal, Germany \\
\noindent
{$^\dagger$~deceased}
    \label{WUPPERTAL}}
\addtolength{\textheight}{-10mm}
\addtolength{\footskip}{5mm}
\clearpage
\headsep 30.0pt
\end{titlepage}
%
\pagenumbering{arabic} 
\setcounter{footnote}{0} %
\large
\section{Introduction}

The first evidence~\cite{vanc} for doubly resonant production of $Z$ bosons
was observed during 1997,
when the LEP-2 accelerator reached a centre-of-mass energy near 183 GeV, corresponding 
to the threshold for this channel. In this article we present measurements of 
the production cross-section, using data collected by the DELPHI
experiment in 1997-2000, at centre-of-mass energies up to 209 GeV, corresponding 
to an integrated luminosity of about 665~pb$^{-1}$.

There are several motivations for studying this channel. Firstly, it enables
a check of the Standard Model (SM) prediction. Both the cross-section and angular 
distribution of the produced $Z$ bosons are sensitive to contributions from new 
physics beyond the SM. Several mechanisms for anomalous production~\cite{hdoublet,grav} 
can hence be constrained directly. For instance, limits on neutral triple gauge boson 
couplings~\cite{hagiwara}, which are forbidden in the SM, can be set~\cite{LEPEW-EP}. 
Secondly, the $ZZ$ production process is an almost irreducible background to the Higgs search
at LEP, particularly when the mass of the Higgs boson is close to that of the 
$Z$~\cite{higgspre}. It provides an environment similar to that corresponding to a 
possible Higgs signal, in the main final state topologies which are analysed, both in 
terms of experimental signature and rate. In this context, it allows to test the 
techniques used in the Higgs analyses on an existing process. 

In what follows, the data sets and simulations used are described and
the signal definition adopted is discussed. The event selections 
are presented for each of the seven sub-channels which were analysed:
$q \bar{q} q \bar{q}$,
$\nu \bar{\nu} q \bar{q}$, 
$\mu^+ \mu^- q \bar{q}$, 
$e^+e^- q \bar{q}$, 
$\tau^+ \tau^- q \bar{q}$, 
$l^+l^-l^+l^-$, and
$\nu \bar{\nu} l^+l^-$ 
(with $l=e,\mu$).
Results are given in the form of a comparison of the numbers of observed and 
predicted selected events, together with an evaluation of the main systematic 
effects. Finally, combinations of sub-channel results into overall $ZZ$ 
cross-sections at each centre-of-mass energy are described, as well as
combinations performed over all centre-of-mass energies, both separately 
in each sub-channel, and for all channels together. The resulting measurements 
are compared with SM expectations.

These results update and supersede those already published at 183 and 
189 GeV~\cite{delphizz189}, by including the $\tau^+ \tau^- q \bar{q}$
sub-channel and using improved methods in several of the other channels. Measurements 
of on-shell $ZZ$ production published by the three other LEP collaborations can 
be found in~\cite{vanc,alephzz,l3zz,opalzz}.

\section{Data samples and simulation}

The data samples used were collected by DELPHI in the years 1997-2000.
The corresponding centre-of-mass energies and integrated luminosities 
which were analysed are given in Table \ref{table:ecmslumi}. 
A detailed description of the detector and a review of its performance can 
be found in~\cite{DELPHIDET,DELPHIPER}. For LEP-2 operation, the vertex detector
was upgraded~\cite{VD} and a set of scintillator counters was added to veto 
photons in blind regions of the electromagnetic calorimetry, at polar angles 
$\theta \simeq 40^\circ$, $90^\circ$ and $140^\circ$.

Simulated events were produced with the DELPHI simulation program 
{\tt DELSIM}~\cite{DELPHIPER} and were then passed through the same
reconstruction and analysis chain as the data. 
The generation of processes leading to four-fermion final 
states was done with {\tt EXCALIBUR}~\cite{EXCALIBUR}, relying on 
{\tt JETSET} 7.4 \cite{JETSET} for quark fragmentation. 
{\tt GRC4F}~\cite{GRC4F} was used as a complementary generator for 
four-fermion final states resulting from single-resonant
$W e \nu_e$ and $(Z/\gamma^*) e^+e^-$ processes when the spectator
electron was close to the beam direction. 
Two-fermion processes $e^+ e^- \rightarrow\ q \bar{q} (\gamma)$
were generated using {\tt PYTHIA} \cite{JETSET}, 
$e^+ e^- \rightarrow\ \mu^+ \mu^- (\gamma)$ and
$e^+ e^- \rightarrow\ \tau^+ \tau^- (\gamma)$ with 
{\tt KORALZ}~\cite{KORALZ},
and $e^+ e^- \rightarrow\ e^+ e^- (\gamma)$ with 
{\tt BHWIDE}~\cite{BHWIDE}. Two-photon interactions were generated 
using {\tt TWOGAM}~\cite{TWOGAM} and {\tt BDK}~\cite{BDK}.
The sizes of the simulated event samples were typically at least
100 times larger than expected in real data.

One sector (corresponding to 1/12) of the main tracking device, the 
Time Projection Chamber, was inactive for the last quarter of the 2000 data 
sample. The corresponding small change of analysis sensitivity from this
period was studied with specially made simulations and taken into account 
in the extraction of the cross-sections.

\begin{table}[htb]
\begin{center}
\begin{tabular}{|c|c|c|} 
\hline
Year       & $\sqrt s$     & Integrated              \\
           & $[$GeV$]$     & luminosity [pb$^{-1}$]  \\ 
\hline
1997       & 182.6         & 54.0                   \\
1998       & 188.6         & 158.1                  \\
1999       & 191.6         & 25.8                   \\
1999       & 195.5         & 76.9                   \\
1999       & 199.5         & 84.3                   \\
1999       & 201.6         & 41.1                   \\
2000       & $<$ 205.5     & 83.3                   \\ 
2000       & $>$ 205.5     & 141.8                  \\
\hline
Total      & -             & 665.3                  \\
\hline
\end{tabular}
\end{center}
\caption{\label{table:ecmslumi} 
Centre-of-mass energies and integrated luminosities of the analysed data. 
During the year 2000, the energies reached were in the range 202-209 GeV, 
clustered mainly around 205 and 207 GeV.
}
\end{table}

\section{Signal definition in the simulation}

  \begin{figure}[!ht]
  \begin{center}
     \SetWidth{1.2}
     \begin{picture}(400,100)(0,100)

     \ArrowLine(40,180)(75,160)
     \ArrowLine(75,160)(75,120)
     \ArrowLine(75,120)(40,100)
     \Photon(75,160)(110,180){3}{6}
     \Photon(75,120)(110,100){3}{6}
     \Vertex(75,160){2}
     \Vertex(75,120){2}
     \Text(40,100)[rb]{\Large e$^{+}$}
     \Text(40,180)[rb]{\Large e$^{-}$}
     \Text(115,180)[lb]{\Large Z}
     \Text(115,100)[lb]{\Large Z}
     \Text(70,140)[r]{\Large e}

     \ArrowLine(280,180)(315,160)
     \ArrowLine(315,160)(315,120)
     \ArrowLine(315,120)(280,100)
     \Photon(315,160)(350,100){3}{8}
     \Photon(315,120)(350,180){3}{8}
     \Vertex(315,160){2}
     \Vertex(315,120){2}
     \Text(280,100)[rb]{\Large e$^{+}$}
     \Text(280,180)[rb]{\Large e$^{-}$}
     \Text(355,180)[lb]{\Large Z}
     \Text(355,100)[lb]{\Large Z}
     \Text(310,140)[r]{\Large e}

  \end{picture}
  \vspace{1em}
  \caption{The Feynman graphs for on-shell $ZZ$ production (referred to as the NC02 graphs)}
  \label{fig:fnc02}
  \end{center}
  \end{figure}
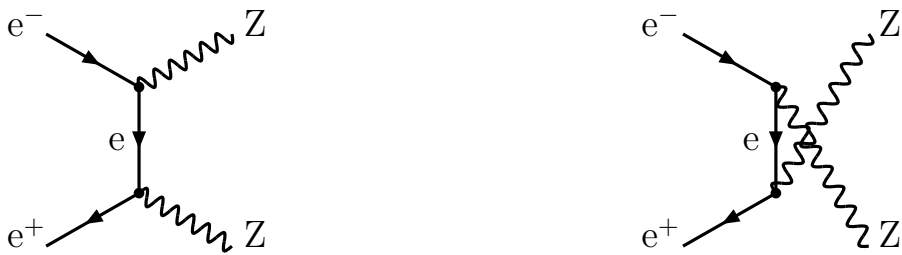

The region of phase-space at high di-fermion masses which characterizes $ZZ$ 
production must be isolated to measure the corresponding cross-section. In this
region, besides the dominant tree-level doubly-resonant $ZZ$ production graphs 
shown in Figure~\ref{fig:fnc02} (referred to as the NC02 graphs), there are also
contributions from other four-fermion processes, which can lead to identical final 
states\footnote{Such contributions arise from $e^+e^- \rightarrow (Z/\gamma^*)\gamma^*$ 
processes (when $\gamma^*$ virtualities are close to the $Z$ boson mass), from 
$(Z/\gamma^*) e^+ e^-$ processes (in cases of final states with electrons), 
and from processes involving $W$ bosons (in cases of $u \bar{u} d \bar{d}$, $c \bar{c} 
s \bar{s}$, and $\nu_l \bar{\nu_l} l^+ l^-$ final states).}. In order to interpret
the measurements in terms of NC02 graphs, the $ZZ$ signal was defined in each of 
the sub-channels by pre-selecting a sub-set of events within the four-fermion simulation,
having both a high purity and efficiency in terms of the relative contribution from the
dominant NC02 graphs. In the analyses of $l^+ l^- q \bar{q}$ (with $l=e,\mu,\tau$) 
final states and in one of the two analyses of the $\nu \bar{\nu} q \bar{q}$ final state
(the probabilistic selection) described in Sections 5-7, the signal definition used was 
based on cuts on the generated boson masses, which were required to be within 10 GeV/c$^2$ 
of the nominal $Z$ mass~\cite{delphizz189}. In the other sub-channels, the signal was 
defined as the sub-set of events satisfying:

\begin{center}
${ {\left|\cal{M_{\rm NC02}} \right|}^2 \over {\left|\cal{M_{\rm All}} \right|}^2 } > 0.5, $
\end{center}

\noindent where $\cal{M_{\rm NC02}} $ and $\cal{M_{\rm All}} $ are the values of the
matrix elements, computed using the four-momenta of the generated fermions, for NC02 and 
for all four-fermion graphs, respectively. Corrections were necessary to relate the rates 
of such events to pure NC02 cross-sections\footnote{Destructive interference from Fermi 
correlations occurs in the case of final states with four identical fermions~\cite{ws.gene}. 
Although modelled by the four-fermion simulation, the resulting reductions in overall 
cross-section, smaller than 0.5\%, were neglected in the calculation of these NC02 
cross-sections.}, in order to take into account the 
efficiency of the signal-defining selection cut, the residual contamination from the 
non-NC02 processes, and interference effects. However, because NC02 and non-NC02 contributions 
are naturally well separated in phase-space, the magnitudes of these corrections were 
typically less than a few percent.
Simulated four-fermion events which did not satisfy the signal-defining cut were considered
as background in the comparison with data.

%
%
%
\section{Four jet channel}
%
%
%

The $q \bar{q} q \bar{q}$ decay mode represents 48.9\% of the expected $ZZ$ 
final states. It results typically in four or more well separated jets 
of particles. The dominant backgrounds originate from fully hadronic 
$WW$ final states and from $q {\bar q} (\gamma)$ processes with hard 
final state gluon radiation, both of which can lead to similar multi-jet 
topologies. The main ingredients required to isolate the signal were di-jet 
mass reconstruction, topological information quantifying the jet separation 
and kinematics and (for $b \bar{b} q \bar{q}$ decay modes) the $b$-tagging 
of the jets. A probabilistic method based on likelihood ratio products was 
developed to combine in an optimal way the information and compute an 
event-by-event measure of the compatibility with the $ZZ$ hypothesis, taking 
into account the flavour content (with or without $b$-quarks) of the signal.

\subsection{Event pre-selection}

A pre-selection was applied to select fully hadronic events. Firstly, events were 
required to contain at least 18 reconstructed tracks of charged particles and to 
have more than 69\% of the available centre-of-mass energy observed in the detector. 
To reduce the contamination from $q {\bar q} (\gamma)$ processes with energetic 
photons emitted in the beam pipe, the effective centre-of-mass energy of the
event, $\sqrt{s'}$, computed as described in \cite{sprime}, was required to 
exceed 80\% of $\sqrt{s}$. Particles were then clustered into jets using the 
DURHAM clustering algorithm \cite{durham} (with $y_{\rm cut}$=0.001) and events 
were selected if at least four jets were reconstructed. 
To reduce backgrounds from processes with energetic and isolated leptons or 
photons, jets were required to contain at least four particles and have an 
invariant mass above 1 GeV/c$^2$. The efficiency of these cuts was about 
88\% and 80\% for the fully hadronic $ZZ$ signal and $WW$ background, respectively,
while about 3\% of events from $q {\bar q} (\gamma)$ processes remained. 
At this stage, backgrounds from all other processes were negligible.

\subsection{Probabilistic selection}

The method developed enabled the computation of the relative probability for any 
pre-selected event to originate from each of three SM processes, $WW$, $ZZ$ or 
$q {\bar q} (\gamma)$. For this purpose a calculation based on likelihood ratio 
products was used, based on the main features characterising the $ZZ$ signal: 
reconstructed di-jet masses, jet kinematics and separation and possible presence of
$b$-hadrons in the jets. These three ingredients and their combination are described below. 
The computation combined measurements of the events with the theoretical expectations 
corresponding to each hypothesis, described analytically where possible in order 
to reduce the dependance on detailed simulations. Throughout the analysis events with 
four jets and events with five or more jets\footnote{Events with more than five jets were 
forced into a five-jet configuration.} were treated separately.

\begin{figure}[h]
\begin{center}
    \mbox{\epsfysize=7.8cm\epsffile{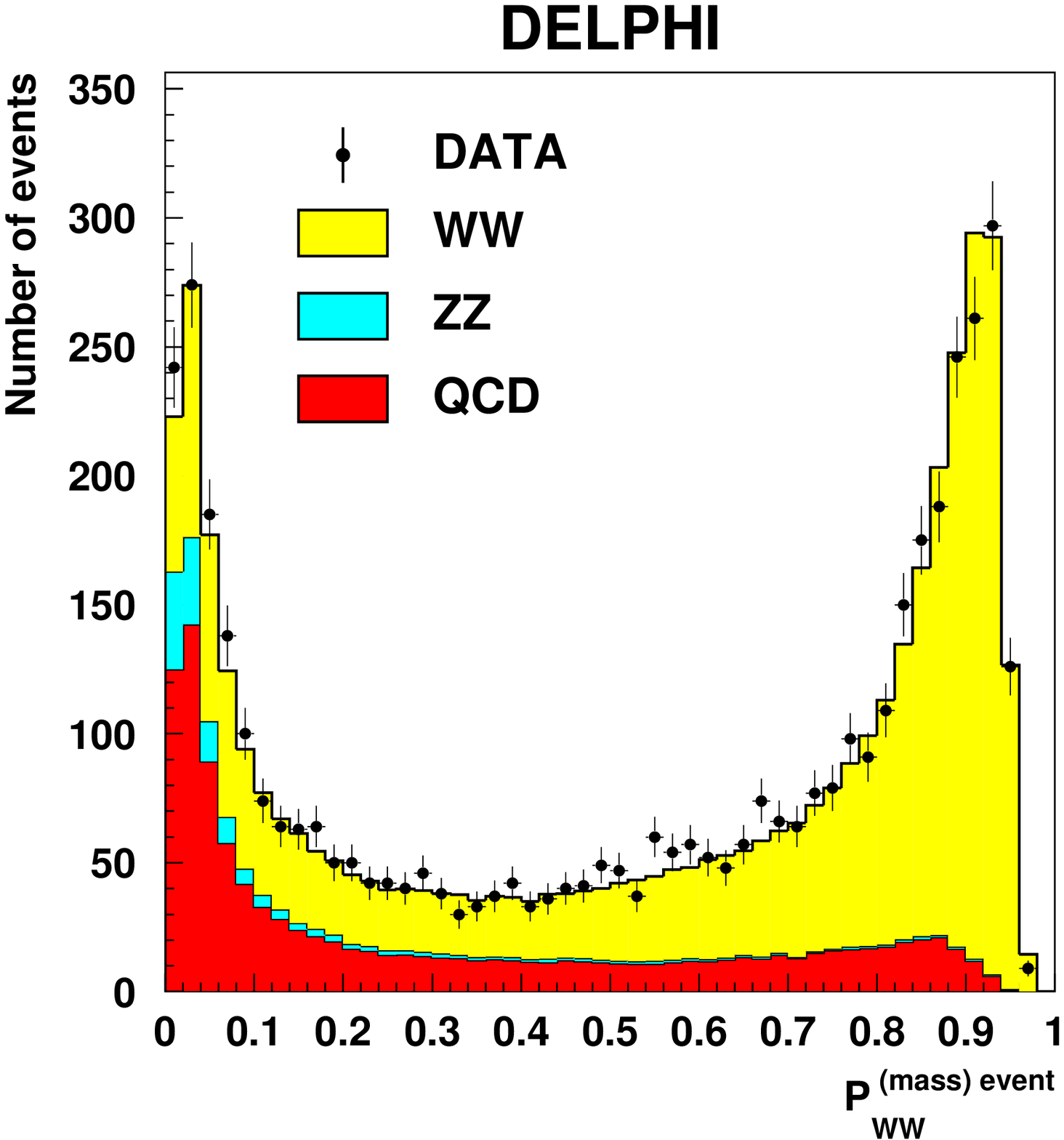}}
    \mbox{\epsfysize=7.8cm\epsffile{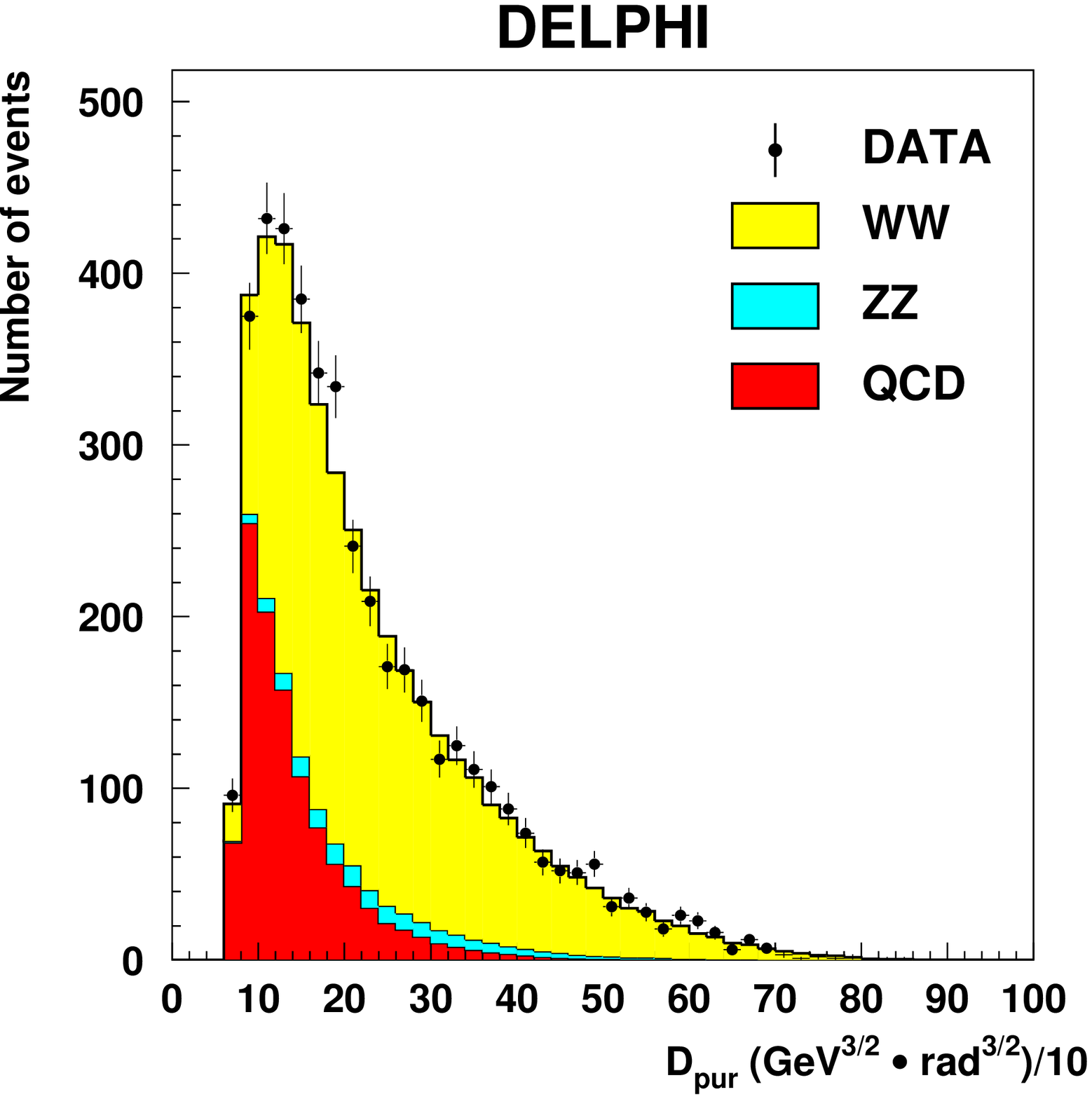}}
\caption{The left-hand plot shows the distribution of the variable quantifying  
for each event the compatibility with the $WW$ hypothesis, computed using only mass 
information. The right-hand plot shows the distribution of the event-topology 
variable D$_{\rm pur}$ (for events with D$_{\rm pur}/10$ greater than 7.5 
GeV$^{3/2}$rad$^{3/2}$). Both distributions are for pre-selected events with four jets
(corresponding to about 80\% of the total sample). The dots are the 
data taken in 1997-2000 and the histograms represent the simulation.
}
\label{fig:massandtopologyinformation}
\end{center}
\end{figure}

Reconstructed mass distributions can be used to characterise the different
hypotheses. The expected shapes are well defined, but in reconstructing events 
ambiguities arise, since in a four or five-jet event, there are three or ten possible 
jet-jet pairings, respectively. In addition, the quality of the energy flow 
reconstruction can vary significantly from event to event. Following the scheme 
described in \cite{ww183}, these considerations were taken into account by 
simultaneously evaluating the compatibility of each event with each hypothesis 
considered, $WW$, $ZZ$ and $q {\bar q} (\gamma)$, from the mass information.
For this purpose, probability distributions for the different mass combinations 
were first expressed in analytical form, as products of Breit-Wigner distributions 
and phase-space factors to describe the case of correctly paired mass combinations in 
genuine $WW$ and $ZZ$ events, and as flat spectra for all the other cases. Then, for each 
reconstructed event, the four-momenta of the jets and their estimated errors were 
used in kinematic fits~\cite{kinfit}, requiring four-momentum conservation and equality 
of the masses of the two jet systems\footnote{Two di-jets in the case of a four-jet 
event, and both a di-jet and a three-jet system in the case of a five-jet event.} with the pairs 
of values to be tested. The two-dimensional distributions of the $\chi^2$ probabilities 
obtained from these fits as a function of the pair of test values were then multiplied 
by the expected probability distributions and the results integrated 
to quantify the compatibility of each pairing with each 
of the $WW$, $ZZ$ and $q {\bar q} (\gamma)$ hypotheses. To compute global event-level 
compatibilities with each hypothesis from the mass information, these results were then 
summed over the possible pairings (assuming equal {\it a priori} weights) and normalised. 
As an illustration, the result obtained for the $WW$ hypothesis is shown in the left-hand 
plot of Figure \ref{fig:massandtopologyinformation}, for the pre-selected events 
with four jets. 
In this method, the inherent 
ambiguity in the pairing of the jets, one of the main difficulties in experimental 
analyses of multi-jet events, did not need to be resolved. 

\begin{figure}[h]
\begin{center}
    \mbox{\epsfysize=7.8cm\epsffile{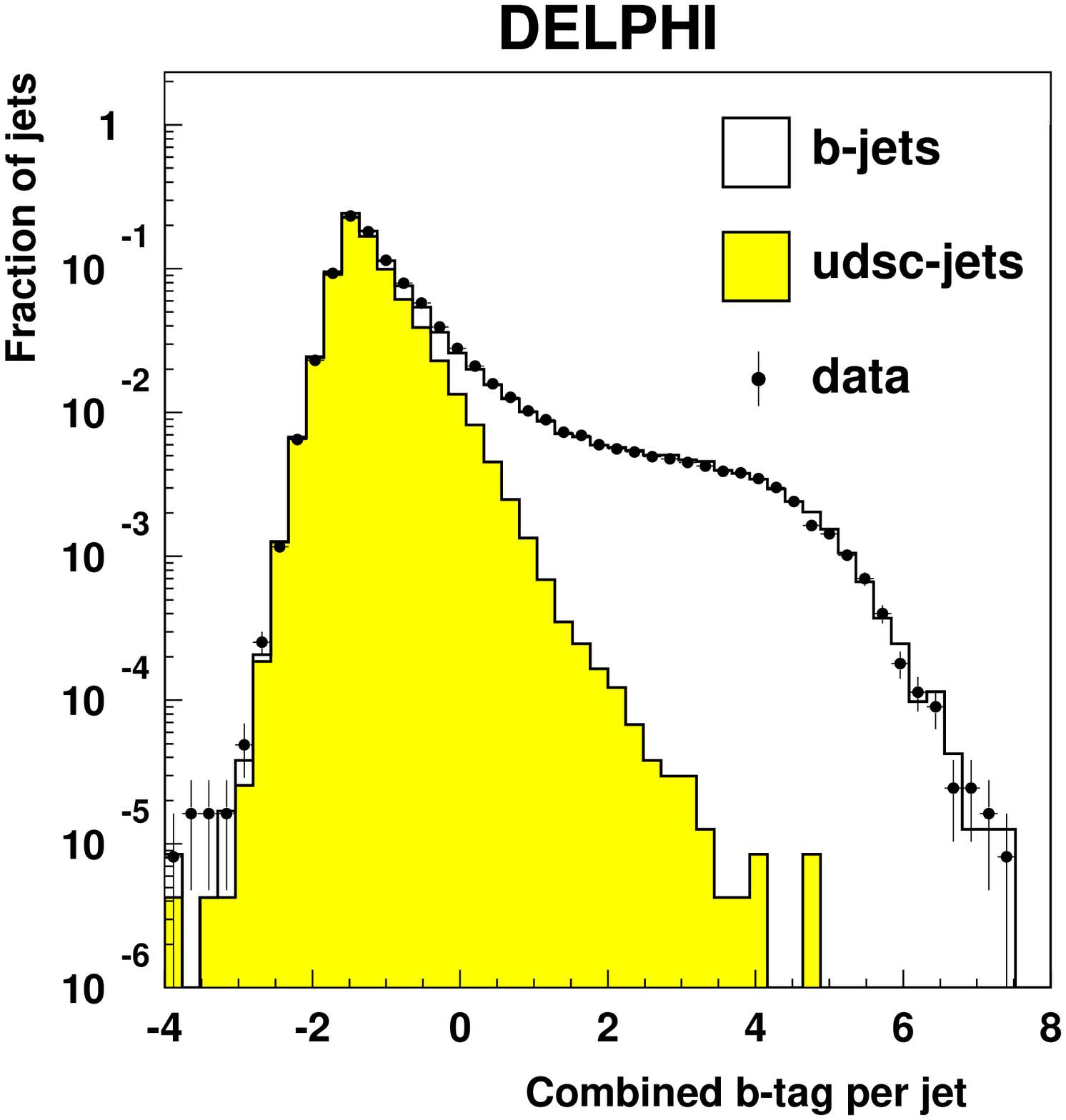}}
    \mbox{\epsfysize=7.8cm\epsffile{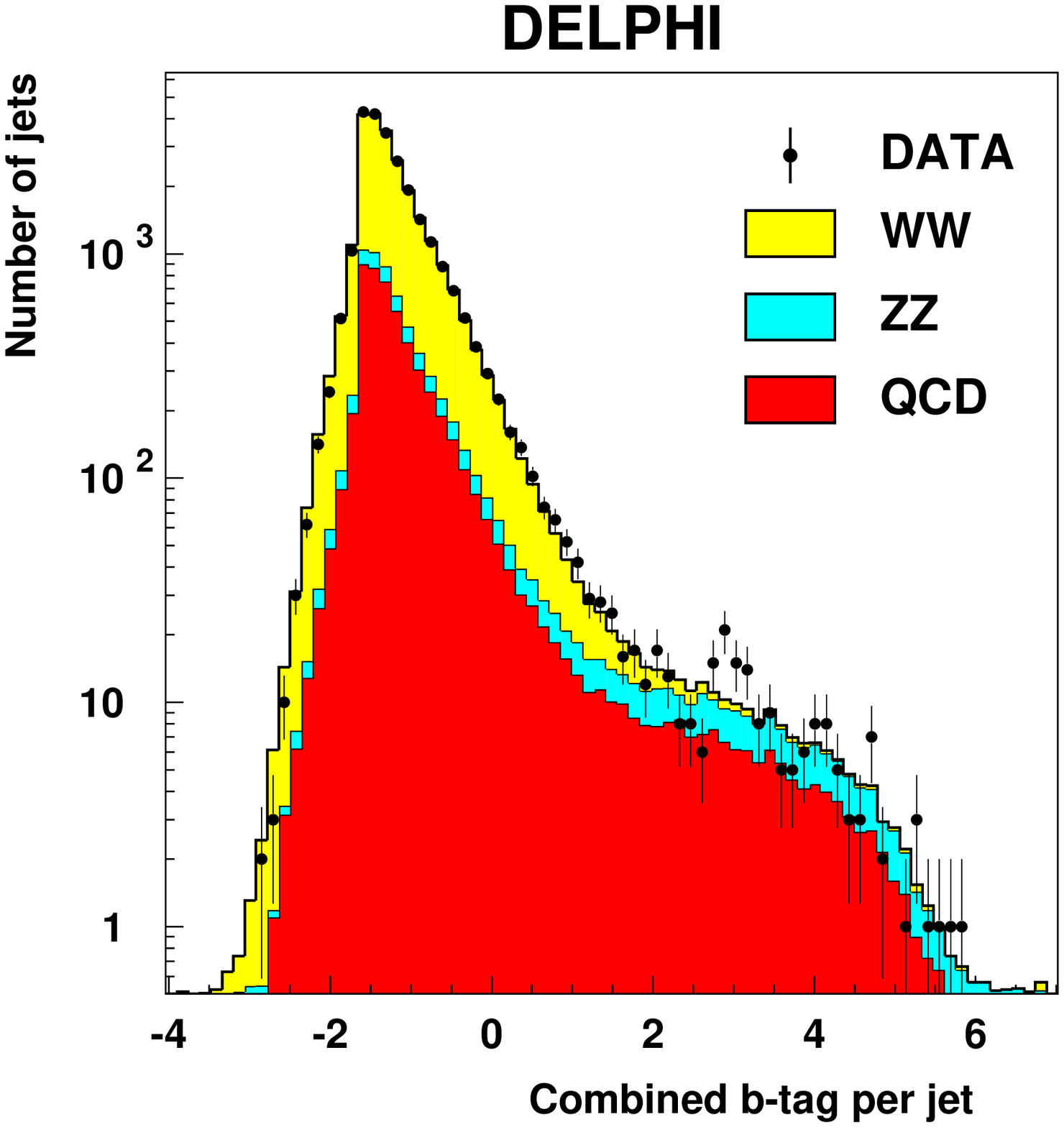}}
\caption{The left-hand plot shows the distribution of the combined $b$-tagging 
variable for jets in $Z$ decays collected in special calibration runs during the 
high-energy data taking. The right-hand plot shows the distribution of the
combined $b$-tagging probability per jet for all reconstructed jets after applying the 
pre-selection. The dots are the data taken in 1997-2000 and the 
histograms represent the simulation.
}
\label{fig:btag}
\end{center}
\end{figure}

Specific features of the jet kinematics and separation in events arising from four-fermion 
and $q {\bar q} (\gamma)$ processes with hard final state gluon radiation leading to a 
four-jet event were also used. Because gluon emission off quarks has infrared 
and collinear divergences, these types of events tend to have cigar-like shapes, while 
the four-fermion processes are more spherical. This difference was exploited to construct a 
topological variable: ${\rm D}_{\rm pur} = E_1 \theta_1 \sqrt{E_2 \theta_2}$, where 
$E_{1}~(E_{2})$ is the leading (subleading) jet energy and $\theta_{1}~(\theta_{2})$ 
is the smallest (next to smallest) opening angle between two jets. The corresponding 
ratio of probability density functions for these
two event types was parametrized as a function of this variable using simulated data 
for four and five jet events separately. The distribution of the ${\rm D}_{\rm pur}$
variable for four-jet events is shown in the right-hand plot of Figure 
\ref{fig:massandtopologyinformation}.

The presence of $b$-hadrons in the jets was identified using a dedicated algorithm~\cite{btag}
which exploited their characteristically long lifetime and high mass. This was 
made possible by the capabilities of the vertex detector~\cite{VD}, in which tracks 
originating from the displaced $b$-hadron vertices could be resolved. The combined 
$b$-tagging variable which was constructed is shown in Figure \ref{fig:btag} for jets 
in hadronic $Z$ decays (left), and in the pre-selected fully hadronic high-energy 
event sample (right). It was particularly powerful in reducing the contamination 
from $WW$ events, since the corresponding final states hardly ever contain a $b$-quark, 
while 38.7\% of fully hadronic $ZZ$ events have at least two $b$-quarks. In order to 
use the information optimally, the ratio of the distributions of this variable for jets 
originating from the fragmentation of $b$ and non-$b$ quarks was parametrized in three 
different angular regions, using simulated $Z$ decays, to take into account the polar 
angle dependence of the detector resolution. Using these parametrizations and the values 
measured for each jet, relative $b$ and non-$b$ probabilities could be assigned to each jet. 

\begin{figure}[h]
\begin{center}
    \mbox{\epsfysize=11.5cm\epsffile{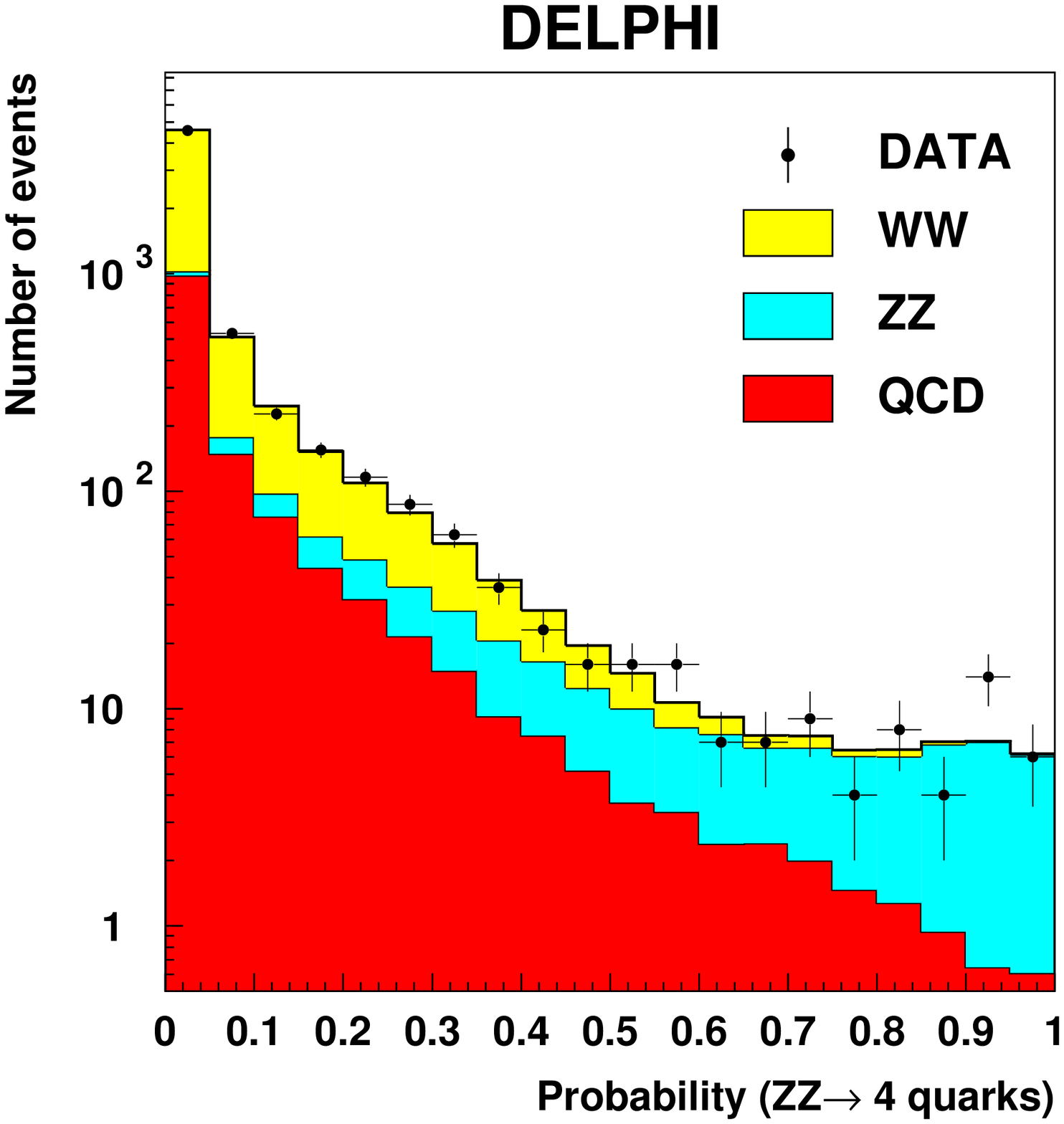}}
\caption{Distribution of the combined $ZZ$ probability. 
The dots are the data taken in 1997-2000 and the  
histograms represent the simulation.
}
\label{fig:qqqqpzzlep2}
\end{center}
\end{figure}

Finally a variable quantifying the compatibility of each event with the $ZZ$ hypothesis 
was constructed using a method based on likelihood ratio products to combine the 
information from the different ingredients described. Its distribution 
is shown in Figure \ref{fig:qqqqpzzlep2}, including the data from all centre-of-mass 
energies. The different energies could be added without diluting the information because
the distributions of the purity of $ZZ$ events with respect to the constructed variable 
were the same.

\begin{table}[htp]
\begin{center}
\begin{tabular}
{|c|c|c|c|c|c|}
\hline
\multicolumn{6}{|c|}{ \ \ \ } \\
\multicolumn{6}{|c|}{ $ZZ \rightarrow q {\bar q} q {\bar q}$ } \\
\multicolumn{6}{|c|}{ \ \ \ } \\
\hline
$\sqrt s$   & Integrated              & Selection       & Predicted  & Predicted & Selected  \\ 
$[$GeV$]$   & luminosity [pb$^{-1}$]  & efficiency      & background & total MC  & data      \\ 
\hline
183         &  54.7                   &   0.13          & 0.92       &    1.84   &    2 \\ 
189         & 158.0                   &   0.27          & 17.88      &   31.13   &   29 \\ 
192         &  25.9                   &   0.34          &  5.31      &    8.64   &   11 \\ 
196         &  76.9                   &   0.38          & 21.16      &   34.01   &   46 \\ 
200         &  84.3                   &   0.39          & 26.64      &   42.09   &   36 \\ 
202         &  41.1                   &   0.42          & 14.24      &   22.55   &   26 \\ 
205         &  82.0                   &   0.41          & 27.45      &   43.73   &   45 \\ 
207         & 142.2                   &   0.42          & 52.58      &   84.13   &   78 \\ 
\hline                                                               
Total       & 665.1                   &   0.38          & 166.18     &  268.11   &  273 \\ 
\hline
\end{tabular}
\end{center}
\caption[.]{
\label{table:qqqqperformance}
Integrated luminosities, selection efficiencies and number of observed and expected 
events, after a cut on the probability of the $ZZ$ hypothesis maximising the product 
of efficiency and purity. The measured $ZZ$ production cross-sections were extracted 
at each centre-of-mass energy by fitting the distributions of this 
probability (see the text).
}
\end{table}

\subsection{Results}

As an illustration, the observed and predicted numbers of events selected after a cut on 
the $ZZ$ probability which maximises the product of efficiency and purity are shown in 
Table \ref{table:qqqqperformance}. There was an overall agreement between data and 
simulation within the statistical fluctuations. At each centre-of-mass energy, a 
measurement of the production cross-section was obtained from a binned maximum likelihood 
fit to the distributions of the $ZZ$ probability, with the $ZZ$ signal contribution as
the only free parameter. The results obtained and the combinations performed between
energies and with other channels to derive global values for the NC02 cross-section 
are described in Section 10. The optimal region of the $ZZ$ probability distribution 
to be used in the fit was determined by minimising the combined statistical and 
systematic uncertainty, evaluated using the full LEP-2 data sample. This resulted in 
a lower cut on the $ZZ$ probability of 0.25. At this level, the number of events from  
the $WW$, $q {\bar q} (\gamma)$ and $ZZ$ processes were comparable.

\subsection{Systematic uncertainties}
 
The main source of systematic error in the selection of $ZZ \rightarrow q \bar {q} q \bar{q}$ 
events was from the limited precision available in the 
modelling of the multi-jet $q {\bar q} (\gamma)$ processes (particularly those involving 
$b$-quarks) which comprised the main background in the most signal-like regions relevant 
to the extraction of $ZZ$ cross-sections. Theoretical uncertainties in the predictions 
and biases from the generator treatments were studied at $\sqrt s = M_Z$, where $M_Z$ is 
the mass of the $Z$ boson, and extrapolated 
to LEP-2 energies~\cite{qcdb}. The most relevant aspects for the measurement of the 
$q {\bar q} q {\bar q} $ channel are summarised below.

It was found that {\tt PYTHIA} \cite{JETSET} underestimated the inclusive four-jet rate\footnote{
The agreement between data and simulation for general hadronic event properties and shapes 
depends on the strategy applied in the tuning~\cite{qcd.tuning} of the generator 
parameters~\cite{qcdb}.} by typically about 10\% at $\sqrt s = M_Z$. The sign and 
magnitude of this discrepancy were also confirmed by comparisons at LEP-2 
energies.
After correcting for $1/2$ of the discrepancy, $\pm 1/2$ of it 
was used as a conservative estimate of the uncertainty in the prediction of this rate.

The probability of secondary $c$ and $b$-quark pair production through gluon 
splitting processes was found to be underestimated in {\tt PYTHIA} 
by factors 
of about 1.5 and 2, respectively, compared to other calculations and dedicated 
measurements.
An approximate correction of this deficit was achieved in the analysis by reweighting 
simulated events containing
gluons splitting into heavy quarks with these factors. Since the gluon splitting process
remains poorly known both theoretically and experimentally, and since the reweighting 
procedure applied provided only a rough correction, a $\pm$50\% relative error on these
probabilities was assumed.

The magnitude of the reduction in gluon radiation off $b$-quarks relative to other flavours 
(arising from their heavier mass) was shown to be overestimated in the version of 
{\tt PYTHIA} 
used, compared to analytic calculations and to dedicated 
measurements, typically by as much as the theoretical uncertainty in these 
calculations.
An additional $\pm$4\% relative error in the four-jet rate 
from $q {\bar q} (\gamma)$ processes was conservatively assumed to cover both this 
uncertainty and bias.

The impact of propagating these uncertainties in the background level to the fitted 
$ZZ \rightarrow q {\bar q} q {\bar q}$ cross-sections is shown in Table~\ref{tab:sys_qqqq}, 
using the full LEP-2 data sample. The effect from uncertainties in selection efficiencies 
related to the $b$-tagging procedure is also shown, as well as the effect from varying the 
$WW$ cross-section within an estimated uncertainty of $\pm$2\%~\cite{ws.gene}. The estimate 
of the error from $b$-tagging was obtained by propagating the errors evaluated 
in~\cite{btag.rb} to the computed $b$ and non-$b$ jet probabilities. All errors 
were assumed to be fully correlated between the centre-of-mass energies analysed.

\begin{table}[htb]
\begin{center}
{\small
\begin{tabular}{|c|c|c|c|c|c|}  
\hline 
$\sqrt s$  & Inclusive    & Gluon splitting  & Gluon radiation& $b$-tagging& $WW$          \\
$[$GeV$]$  & four-jet rate& into heavy quarks& off $b$-quarks & procedure  & cross-section \\ 
\hline
183 - 208  & $\pm$2.3 \% & $\pm$2.3 \%     & $\pm$1.8 \%   & $\pm$1.5 \% & $\pm$0.9 \%\\ 
\hline
\end{tabular}
}
\end{center}
\caption{\label{tab:sys_qqqq} 
Expected relative uncertainties on the fitted $ZZ \rightarrow q {\bar q} q {\bar q}$ 
cross-section from systematic uncertainties in the predicted background and selection 
efficiencies.
}
\end{table}

%
%
%
\section{Jets and missing energy}
%
%
%

The $\nu \bar{\nu} q \bar{q}$ decay mode represents 28.0\% of the $ZZ$ final states. 
The event topology is characterised by a pair of acoplanar jets (acollinear in the 
transverse plane to the beam) with visible and recoil masses compatible with the $Z$ 
mass. Because two energetic neutrinos escape detection in this channel, efficient and 
reliable energy flow reconstruction is essential to select the signal. The most 
difficult physical backgrounds arise from single-resonant $W e \nu_e$ processes, 
from $WW$ processes where one $W$ decays into $\tau \nu_{\tau}$, and from 
$q\bar{q}$ events, which may or may not be accompanied by energetic isolated photons 
escaping detection, in which one or both jets are badly reconstructed. 
 
After a common pre-selection, two independent analyses were carried out. Each used a 
separate selection of discriminating variables to construct a combined estimator, 
employing different, complementary schemes. The first analysis used a non-linear 
discriminant method (the so-called Iterative Discriminant Analysis (IDA)~\cite{IDA}) 
in which the cross-products between the variables were included, and where 
the correlations - important in this channel - can be treated optimally. The second 
analysis used a more conventional probabilistic method based on likelihood ratio products,
in which the non-Gaussian profiles of the variables are taken into account. Results from 
both analyses were obtained and compared. The IDA analysis had slightly better performance, 
and was chosen to derive the combined values of the NC02 cross-section described in Section 
10. It was also used to study the propagation of systematic uncertainties. The probabilistic 
analysis served as a cross-check.



\subsection{Event pre-selection}

Selecting carefully the measured particles from which event variables are calculated is 
important to obtain efficient and reliable energy flow reconstruction. The definition of 
neutral particles was studied specially in this context, using both energy clusters in the 
calorimeters not associated to charged particle tracks and reconstructed vertices of photon 
conversions, of interactions of neutral hadrons and of decays of neutral particles in the 
tracking volume. 
Beam related backgrounds were also rejected by requiring at least two charged particles 
with impact parameter (with respect to the fitted primary vertex) less than 1~mm in the 
transverse plane and less than 3~mm along the beam axis, and with transverse momentum, 
$P_t$, greater than 2 GeV/$c$. A loose hadronic pre-selection was then applied, 
requiring at least eight charged particles, total charged energy greater than 
$0.16\sqrt{s}$, transverse energy (defined as $\Sigma P_t$, where the sum is over
all particles)
greater than $0.15\sqrt{s}$, and the sum of the component of 
momentum of all particles along the thrust axis greater than  $0.25\sqrt{s}$. 
Finally, events with an electromagnetic shower of energy 
exceeding $0.45\sqrt{s}$ were rejected. This removed about 97\% of the 
background from two-photon and Bhabha processes.

The particles of the events were forced into two-jet configurations using the DURHAM 
jet algorithm. To reject events coming from $q {\bar q} (\gamma)$ processes with 
energetic photons emitted in the beam pipe, the effective centre-of-mass energy
of the event, $\sqrt{s'}$, computed from the jet directions, was required to be greater 
than 115~GeV when the polar angle of the total momentum of detected particles 
was less than $40^\circ$ or larger 
than $140^\circ$. To reduce the contamination from $q {\bar q} (\gamma)$ 
processes with energetic photons in the detector acceptance, events 
were rejected if their total electromagnetic energy within $30^\circ$ of 
the beam axis was greater than $0.16\sqrt{s}$, or if the total energy 
in the luminosity monitor was greater than $0.08\sqrt{s}$. To reject 
$q {\bar q} (\gamma)$ processes with energetic photons emitted in blind 
regions of the electromagnetic calorimetry (at polar angles 
$\theta \simeq 40^\circ$, $90^\circ$ and $140^\circ$) signals from the 
set of dedicated scintillator counters were used. To reduce the two-fermion 
background without significant initial state radiation as well as four-fermion 
backgrounds without missing energy, the effective centre-of-mass energy of the event 
had to satisfy $\sqrt{s'} < 0.96\sqrt{s}$. 
To reduce backgrounds from two-fermion events with jets pointing 
to the insensitive regions of the electromagnetic calorimeters, where 
the energy flow was less precise, events were rejected if the jets 
had polar angles within the ranges $35-45^\circ$ or $135-145^\circ$, 
unless the acoplanarity of the event was greater than $10^\circ$. This 
removed about 88\% of the total $q {\bar q}(\gamma)$ background.

The contamination from $WW$ processes where one $W$ decayed leptonically 
was reduced by requiring that the energy of the most energetic particle of 
the event be less than $0.2\sqrt{s}$. To improve the rejection when the $W$ 
decayed into $\tau \nu_{\tau}$, the events were required to have no charged 
particle with a transverse momentum with respect to its jet direction 
greater than 10 GeV/$c$. This removed about 66\% of the $WW$ background.

At the end of the event pre-selection, the efficiency for the 
$ZZ \rightarrow \nu \bar{\nu} q \bar{q}$ signal was about 77\%.

\subsection{IDA selection}

\begin{figure}[h]
  \begin{center}
    \mbox{\epsfysize=12.0cm\epsffile{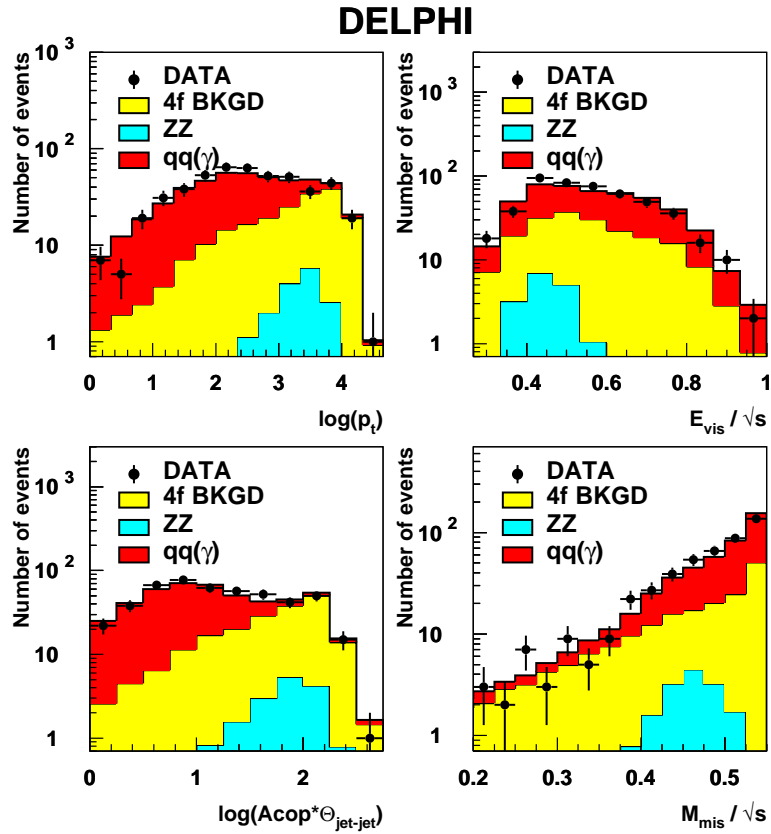}}
    \caption{
       Distributions of four variables used in the IDA analysis stream
       shown after all pre-selection cuts (see text): logarithm of the
       transverse momentum (in GeV/c) with respect to the beam axis (upper
       left), ratio of the visible energy to the centre-of-mass energy
       (upper right), logarithm of the product of the inter-jet angle 
       (in degrees) and the acoplanarity (lower left), ratio of the missing 
       mass (computed using the constraint that the visible mass be compatible 
       with $M_Z$) to the centre-of-mass energy (lower right). 
       The dots are the data taken in 1999 at a centre-of-mass energy 
       $\sqrt s$ = 199.5 GeV. The histograms are the simulation prediction.
       }
    \label{fig:presel}
  \end{center}
\end{figure}

After the common pre-selection described in Section 5.1, it was additionally required 
that the momentum of the most isolated particle in the event be in the range 
$0.01 \sqrt{s} < p < 0.20 \sqrt{s}$, and that the visible mass be smaller than $\sqrt{s}$.
A combined discriminant variable was then constructed using the Iterative 
Discriminant Analysis program (IDA)~\cite{IDA} to calculate a second order 
polynomial from twelve event variables, selected according to their discriminating 
power and independence
\footnote{To concentrate on the signal region when optimizing the second-order discriminant 
function, and to avoid long non-Gaussian tails, additional very loose pre-selection cuts 
were applied to some of the variables, as indicated in the description.}:

\begin{itemize}
\item
The logarithm of the transverse momentum of the event with respect to the beam axis,
for values greater than 0 (see Figure~\ref{fig:presel} upper left),
\item
The visible energy of the event normalised to $\sqrt s$,
for values lower than 1 (see Figure~\ref{fig:presel} upper right),
\item
The transverse energy of the event normalised to $\sqrt s$,
for values in the range $0.15-0.60$,
\item
The minimum polar angle defining a cone in the positive 
and negative beam directions containing 6\% of the total visible energy,
\item
The difference of the jet energies normalised to $\sqrt s$,
for values lower than 0.35,
\item
The sum of energies of particles having angles with respect to the most 
isolated particle larger than $5^\circ$ and smaller than $60^\circ$ 
($25^\circ$ if the momentum of the most isolated particle exceeded 
5 GeV/$c$), normalised to $\sqrt s$, for values lower than 0.15,
\item
The effective centre-of-mass energy $\sqrt{s'}$, 
\item
The logarithm of the product of the acoplanarity 
with the inter-jet angle, for values in the range
$0.0-2.8$ (see Figure~\ref{fig:presel} lower left),
\item
The logarithm of the acollinearity of the two jets,
\item
The longitudinal momentum with respect to 
the thrust axis, normalised to $\sqrt s$, for values 
lower than 1,
\item
The missing mass normalised to $\sqrt s$, computed 
with the constraint that the visible mass equal $M_Z$ 
(see Figure~\ref{fig:presel} lower right), 
\item
The logarithm of the largest transverse momentum of any particle 
with respect to its jet direction.
\end{itemize}

\begin{figure}[tbh]
  \begin{center}
    \mbox{\epsfysize=7.8cm\epsffile{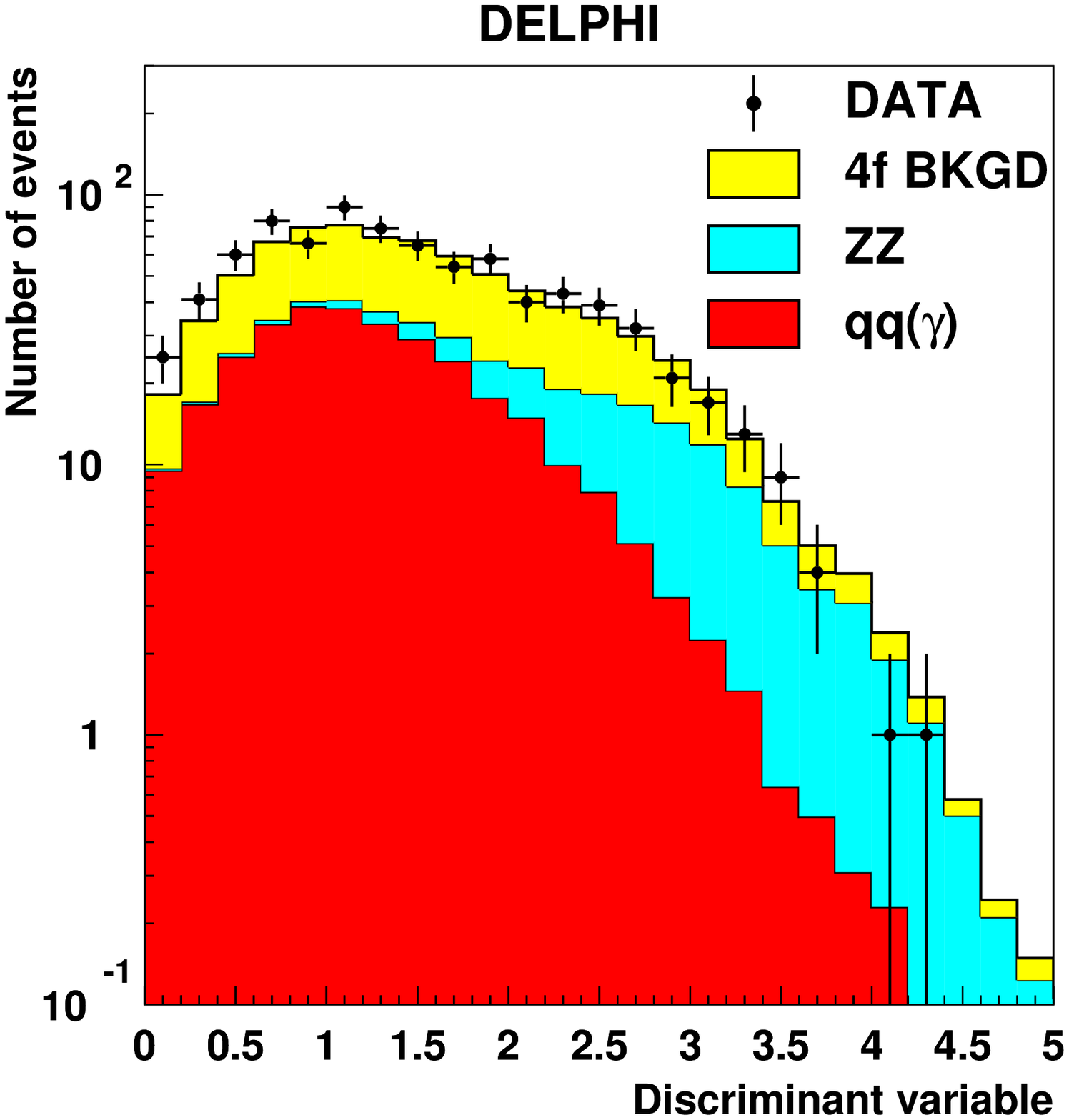}}
    \mbox{\epsfysize=7.8cm\epsffile{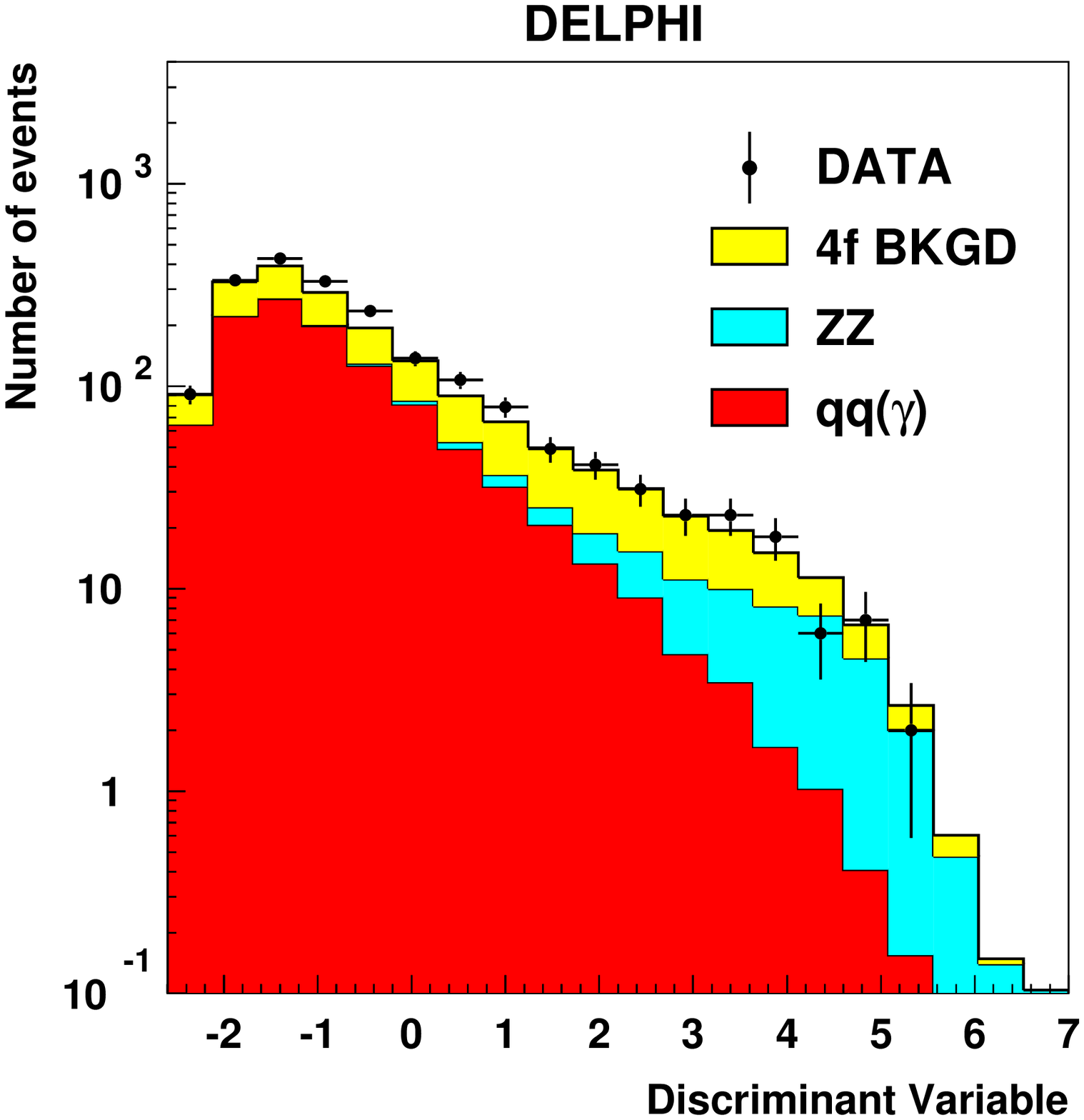}}
    \caption{
      Distribution of the combined discriminating variable for the
      $\nu \bar{\nu} q \bar{q}$ analysis in the IDA (left)
      and probabilistic (right) selections.
      The dots are the data taken in 1997-2000 and
      the histograms represent the simulation.
      The distributions of the purity of $ZZ$ events as a function of these two
      variables were almost identical for all centre-of-mass energies, allowing to add 
      the histograms of the different energies with minimal dilution of the information. 
       }
    \label{fig:qqnncomb}
  \end{center}
\end{figure}

The observed distributions of these twelve variables were in 
good agreement with the predictions from simulation at each centre-of-mass
energy analysed. As an illustration, the distributions of four of them
are shown in Figure~\ref{fig:presel} after all pre-selection cuts, for
$\sqrt{s}=199.5$ GeV. The IDA method for variable combination was applied 
in two successive optimisations. Following the first one, an enriched 
sub-sample of events was selected, with a 95\% relative signal efficiency. 
These events were then used in the second optimisation. The final combined discriminant 
variable obtained is shown in the left-hand plot of Figure~\ref{fig:qqnncomb}, 
including the data from all centre-of-mass energies.

\subsection{Probabilistic selection}

After the common pre-selection described in Section 5.1, it was additionally
required that the total visible energy be lower than $0.70\sqrt{s}$. 
A probabilistic method based on likelihood ratio products was then used 
to combine nine discriminating variables:

\begin{itemize}
\item
The energy of the most energetic jet,
\item
The cosine of the acoplanarity,
\item
The effective centre-of-mass energy $\sqrt{s'}$, normalised to $\sqrt s$, 
\item
The missing momentum,
\item
The cosine of the polar angle of the missing momentum,
\item
The largest transverse momentum of any particle with respect to its jet direction,
\item
The minimum charged multiplicity of the jets\footnote{In this case events were not forced into 
two-jet configurations and the DURHAM jet clustering algorithm was used with $y_{cut}=0.005$.},
\item
The visible mass, computed with the constraint that the 
missing mass equal $M_Z$, 
\item
The output of a veto algorithm based on the response of the 
dedicated scintillator counters installed at polar angles 
$\theta \simeq 40^\circ$, $90^\circ$ and $140^\circ$.
\end{itemize}

There was good agreement between the data and the predictions 
from the simulation for each of these variables at each centre-of-mass 
energy. The probability density functions used to construct 
the combined discriminant variable were determined from the simulation, separately at each 
energy. The distribution of this final variable is shown in the right-hand plot of
Figure~\ref{fig:qqnncomb}, including the data from all centre-of-mass energies.

\subsection{Results and comparison}

The results from the two selections are presented in 
Table~\ref{tab:comp_ida_lh}, where the observed and predicted 
numbers of events, selected after cuts on each discriminant variable 
maximising the product of the efficiency and purity, are 
shown\footnote{The luminosities used to
analyse the $ZZ \rightarrow \nu \bar{\nu} q \bar{q}$ channel
were slightly lower than for other channels, because additional 
criteria on the operation of the apparatus were used.}. 
The two methods gave compatible results, but the IDA 
selection performed slightly better. There was overall agreement 
between data and simulation within statistical errors in both cases.
A study of the overlap between the two selections, performed with the data 
collected in 2000, showed that the proportion of common events was about 
60\% in real data: 55\% in the background and 85\% in the signal, according to
the simulation. The IDA 
selection was used as the main analysis, and the probabilistic as cross-check. 

\begin{table}[htp]
\begin{center}
\begin{tabular}
{|c|c|c|c|c|c|} 
\hline
\multicolumn{6}{|c|}{} \\
\multicolumn{6}{|c|}{ $ZZ \rightarrow \nu \bar{\nu} q \bar{q}$ } \\
\multicolumn{6}{|c|}{ (IDA selection) } \\
\multicolumn{6}{|c|}{} \\ 
\hline 
$\sqrt s$   & Integrated              & Selection       & Predicted  & Predicted & Selected  \\ 
$[$GeV$]$   & luminosity [pb$^{-1}$]  & efficiency      & background & total MC  & data      \\ 
\hline
183         & 52.4   & 0.31    &  1.6   &  3.1   &  2 \\
189         &152.8   & 0.49    & 14.8   & 28.3   & 22 \\
192         & 24.9   & 0.56    &  5.1   &  8.1   &  6 \\
196         & 75.0   & 0.52    & 13.3   & 22.6   & 27 \\
200         & 82.2   & 0.56    & 18.7   & 31.0   & 25 \\
202         & 40.4   & 0.48    &  7.5   & 13.0   & 12 \\
205         & 72.9   & 0.54    & 16.8   & 27.9   & 22 \\
207         &138.4   & 0.52    & 30.7   & 51.9   & 49 \\ 
\hline 
Total       &639.0   & 0.50    &108.5   &185.9   &165 \\ 
\hline
\hline
\multicolumn{6}{|c|}{} \\
\multicolumn{6}{|c|}{ $ZZ \rightarrow \nu \bar{\nu} q \bar{q}$ } \\
\multicolumn{6}{|c|}{ (Probabilistic selection) } \\
\multicolumn{6}{|c|}{} \\ 
\hline 
$\sqrt s$   & Integrated              & Selection       & Predicted  & Predicted & Selected  \\ 
$[$GeV$]$   & luminosity [pb$^{-1}$]  & efficiency      & background & total MC  & data      \\ 
\hline
183         & 52.4   & 0.27    &  1.8   &  3.1   &  4 \\
189         &152.8   & 0.46    & 24.3   & 36.7   & 32 \\
192         & 24.9   & 0.50    &  5.6   &  8.5   &  9 \\
196         & 75.0   & 0.46    & 12.7   & 21.4   & 20 \\
200         & 82.2   & 0.46    & 16.4   & 26.0   & 25 \\
202         & 40.4   & 0.51    & 10.9   & 16.6   & 16 \\
205         & 72.9   & 0.48    & 18.0   & 28.4   & 24 \\
207         &138.4   & 0.45    & 29.6   & 47.6   & 48 \\ 
\hline 
Total       &639.0   & 0.45    &119.3   &188.3   &178 \\ 
\hline
\end{tabular}
\end{center}
\caption{\label{tab:comp_ida_lh} 
Integrated luminosities, selection efficiencies and number of observed and 
expected events in the $ZZ \rightarrow \nu \bar {\nu} q \bar{q}$ channel, 
after cuts on each discriminant variable maximising the product of the 
efficiency and purity. In the IDA analysis, the measured $ZZ$ 
production cross-sections were extracted at each centre-of-mass energy by 
fitting the distributions of the corresponding discriminant variable (see the text).
}
\end{table}

At each centre-of-mass energy, a measurement of the production cross-section 
was obtained from a binned maximum likelihood fit to the distribution of the 
final IDA variable, with the $ZZ$ signal contribution as the only free parameter. 
The results obtained and the combinations performed between energies and with other 
channels to derive global values for the NC02 cross-section are described in Section 
10. The optimal region of the distribution of the IDA variable to be used in the fit 
was determined by minimising the combined statistical and systematic uncertainty, 
evaluated using the full LEP-2 data sample. This resulted in removing events with
values of the IDA variable less than 2.2. The cross-section measured by combining the results of the probabilistic 
cross-check analysis over the different energies is also given in Section 10 for 
comparison, using the numbers of events and efficiencies presented in 
Table~\ref{tab:comp_ida_lh}. 

The remaining background was composed of several processes. Near the threshold for 
$ZZ$ production, the main backgrounds were from $q\bar{q}(\gamma)$ events with badly 
reconstructed jets, and from $WW$ processes. At higher energies, mis-reconstructed 
$q\bar{q}(\gamma)$ events were easier to suppress, due to the Lorentz boost of the $Z$ 
bosons in the signal. In this case, the main background was from single-resonant 
$W e \nu_e$ processes.

\subsection{Systematic errors}

The main sources of systematic error in the selection of
$ZZ \rightarrow \nu \bar {\nu} q \bar{q}$ events were from 
uncertainties in the description of the energy flow reconstruction 
of the $q\bar{q}(\gamma$) background and, to a lesser extent, from
uncertainties in the predicted background cross-sections, particularly 
those of the single-resonant $W e \nu_e$ process. The propagation of 
these uncertainties to the final steps of the IDA analysis was studied.

Large samples of events at the $Z$ peak, collected in the same conditions as the 
high energy data, were used to evaluate the errors in the energy flow reconstruction. 
The selection of $ZZ \rightarrow \nu \bar{\nu} q \bar{q}$ events exploits the large
missing energy characteristic of this channel, and hence is sensitive to 
the description of the low energy tail in the reconstruction of background 
processes such as $q\bar{q}(\gamma$). The $Z$ events were 
compared with the simulation to estimate corrections to the particle flow 
in several momentum bins, separately for the barrel and endcaps, and for 
charged and neutral particles. The corrections, consisting mainly of changes 
in multiplicities to account for observed efficiency losses and possible 
duplication effects in the pattern recognition, were typically only a few 
per cent for charged particles, but could be larger in the case of neutral 
particles at low momentum or in the endcaps. It was found that the simulation 
overestimated particle reconstruction efficiencies in the highest momentum 
bins. This resulted in underestimating the lower tail of the total energy 
distribution. Application of the computed corrections significantly improved 
the overall agreement between data and simulation at the pre-selection level for 
all energy flow observables such as the total charged and neutral energies, the 
visible mass or the acoplanarity. Partial improvement also resulted in the tails 
of the distributions most relevant to the selection of $ZZ \rightarrow \nu \bar{\nu} 
q \bar{q}$ events, although comparisons were in this case limited by the statistics 
of the available samples.

This correction procedure was then applied to all simulated high energy samples 
and the IDA analysis was repeated. The main effect was to increase the 
$q\bar{q}(\gamma$) background in the high purity regions relevant to the 
signal extraction by about one third on the average, while other components 
were much less affected. The $ZZ$ cross-sections were then fitted, at each 
centre-of-mass energy, using these modified versions of the simulation. The 
differences in expected results were always of the same sign, corresponding to 
reductions in the $ZZ$ cross-section obtained. After correcting for $1/2$ of 
these discrepancies, $\pm 1/2$ of them were used as conservative estimates 
of the errors resulting from uncertainties in the energy reconstruction.
The magnitude of the effects decreased with energy, because 
of the decreasing relative importance of the $q\bar{q}(\gamma$) background. 

The impact of propagating these uncertainties in the energy reconstruction to 
the fitted $ZZ \rightarrow q {\bar q} \nu {\bar \nu}$ cross-sections is 
shown in Table~\ref{tab:sys_comb}, using the full LEP-2 data sample. The 
combined effect from uncertainties in background cross-sections is also 
shown, assuming the following errors~\cite{ws.gene}: $WW: \pm$ 2\%, 
$We\nu: \pm$ 5\%, $q\bar{q}(\gamma): \pm$ 2\%. Both types of error were assumed 
to be fully correlated between the energies analysed.

\begin{table}[htb]
\begin{center}
\begin{tabular}{|c|c|c|} 
\hline
$\sqrt s$         & Systematic error from   & Systematic error from  \\
$[$GeV$]$         & energy flow             & MC cross-sections      \\ 
\hline
183               & $\pm$14.5 \%           & $\pm$4.0 \%           \\
189               & $\pm$10.0 \%           & $\pm$2.5 \%           \\
192               & $\pm$7.5  \%           & $\pm$2.5 \%           \\
196 - 208         & $\pm$3.5  \%           & $\pm$2.5 \%           \\ 
\hline
\end{tabular}
\end{center}
\caption{\label{tab:sys_comb} Expected effects on the 
$ZZ \rightarrow \nu \bar {\nu} q \bar{q}$ cross-section 
fitted in the IDA analysis, from uncertainties in the energy 
flow reconstruction and predicted background cross-sections.
}
\end{table}


%
%
%
\section {Jets and a pair of isolated muons or electrons \label{sub:QQLL}}
%
%
%

The $\mu^+ \mu^- q \bar{q}$ and $e^+ e^- q \bar{q}$ decay modes
comprise 9.3\% of the $ZZ$ final states. 
The two final state leptons are typically well isolated from all other 
particles. This feature can be used to select such events with high efficiency 
and low background contamination in both the muon and electron channels. 
Events were selected 
initially
without explicit cuts on the masses of the final state fermion pairs, 
in order to analyse simultaneously the production of $ZZ$ and
$Z\gamma^*$ events and, in the case of the $e^+e^- q \bar{q}$ channel, 
contributions from the $(Z/\gamma^*) e^+e^-$ process. 
Mass cuts were then applied to isolate the $ZZ$ contribution.


\subsection{Selection procedure}

A loose hadronic pre-selection was first applied, requiring that the events 
have at least 7 charged particles and a charged energy above 0.30 $\sqrt{s}$. 
To reduce the radiative return to the $Z$ boson, events were rejected 
if a photon with energy greater than 60 GeV was found. The selection procedures 
were then closely similar for both $\mu^+ \mu^- q \bar {q}$ and $e^+ e^- q \bar {q}$ 
channels. In order to maximize the lepton identification efficiency, any charged particle 
with momentum exceeding 5 GeV/c was considered as a possible lepton candidate,
around which nearby photons, if present, could be clustered\footnote{ 
For this purpose any neutral particle identified as a photon with energy greater 
than 0.5 GeV was combined with the lepton candidate if the invariant mass of
the resulting cluster did not exceed 0.4 GeV/c$^2$. At most two photons were included 
in such a cluster, the photon giving the smallest mass increase being added first.}.
This was found necessary to improve its energy evaluation in the presence of final 
state radiation. In the case of the $e^+ e^- q \bar {q}$ channel, photons with energy 
between 20~GeV and 60~GeV were also considered as possible electron candidates, 
to recover events in which the electron track was not reconstructed.

Events with at least two lepton candidates of the same flavour, opposite
charge (if none of the electron candidates originated from a photon)
and invariant mass exceeding 2 GeV/c$^2$ were then selected. 
All particles except the lepton candidates were clustered into jets 
using the JADE algorithm \cite{lund} (with $y_{min}$ = 0.01)
and a kinematic fit requiring four-momentum conservation was applied,
correcting appropriately the errors on lepton energies in cases where
photons had been added by the clustering procedure.

At least one of the two lepton candidates was required to satisfy 
strong lepton identification criteria, while weaker requirements
were specified for the second. Muons were considered as strongly 
identified if selected by the standard DELPHI muon identification 
package \cite{DELPHIPER}, based mainly on finding associated hits 
in the muon chambers. For weak muon identification only a set of 
kinematic and calorimetric criteria was used.
Electrons were considered as strongly identified when the energy 
deposited in the electromagnetic calorimeter exceeded 15 GeV or 60\% 
of the total energy of particles composing the electron candidate 
cluster and when the energy deposited in the hadron calorimeter 
was consistent with the expectations for electrons.
For weak electron identification only requirements 
on the momentum of the charged particle in the cluster and on the
energy deposited in the hadron calorimeter were used. Moreover electron 
candidates originating from applying the clustering procedure around 
a photon were considered as weakly identified.

Two discriminating variables were then used to select the events: 
$P_t^{min}$, the lesser of the transverse momenta of the lepton candidates 
with respect to their nearest jet, and the $\chi^2$ per degree of freedom 
of the kinematic fit. 

\subsection{Results}

\begin{figure}[h]
  \begin{center}
    \mbox{\epsfysize=6.2cm\epsffile{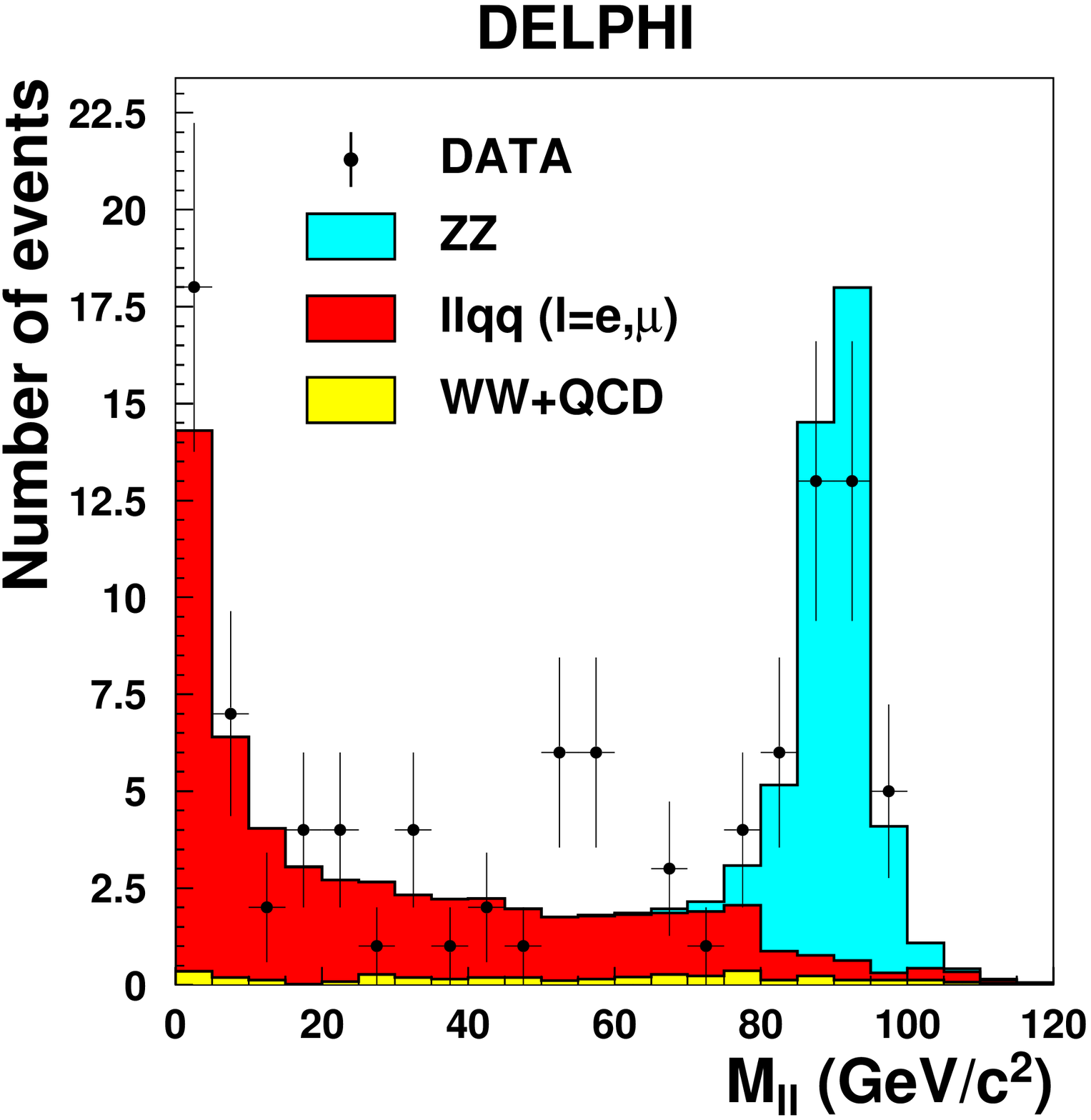}}
    \mbox{\epsfysize=6.2cm\epsffile{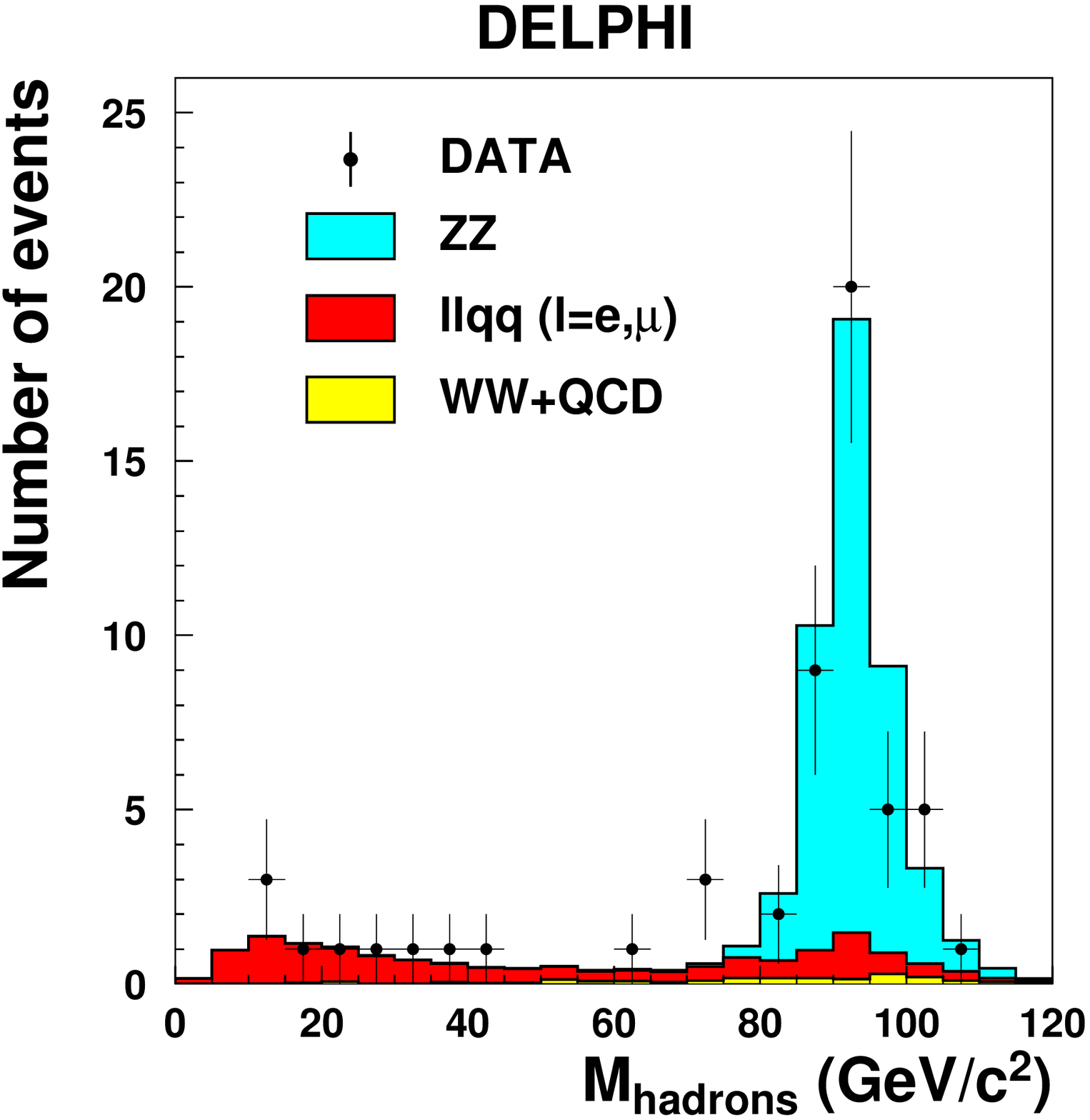}}
    \mbox{\epsfysize=6.2cm\epsffile{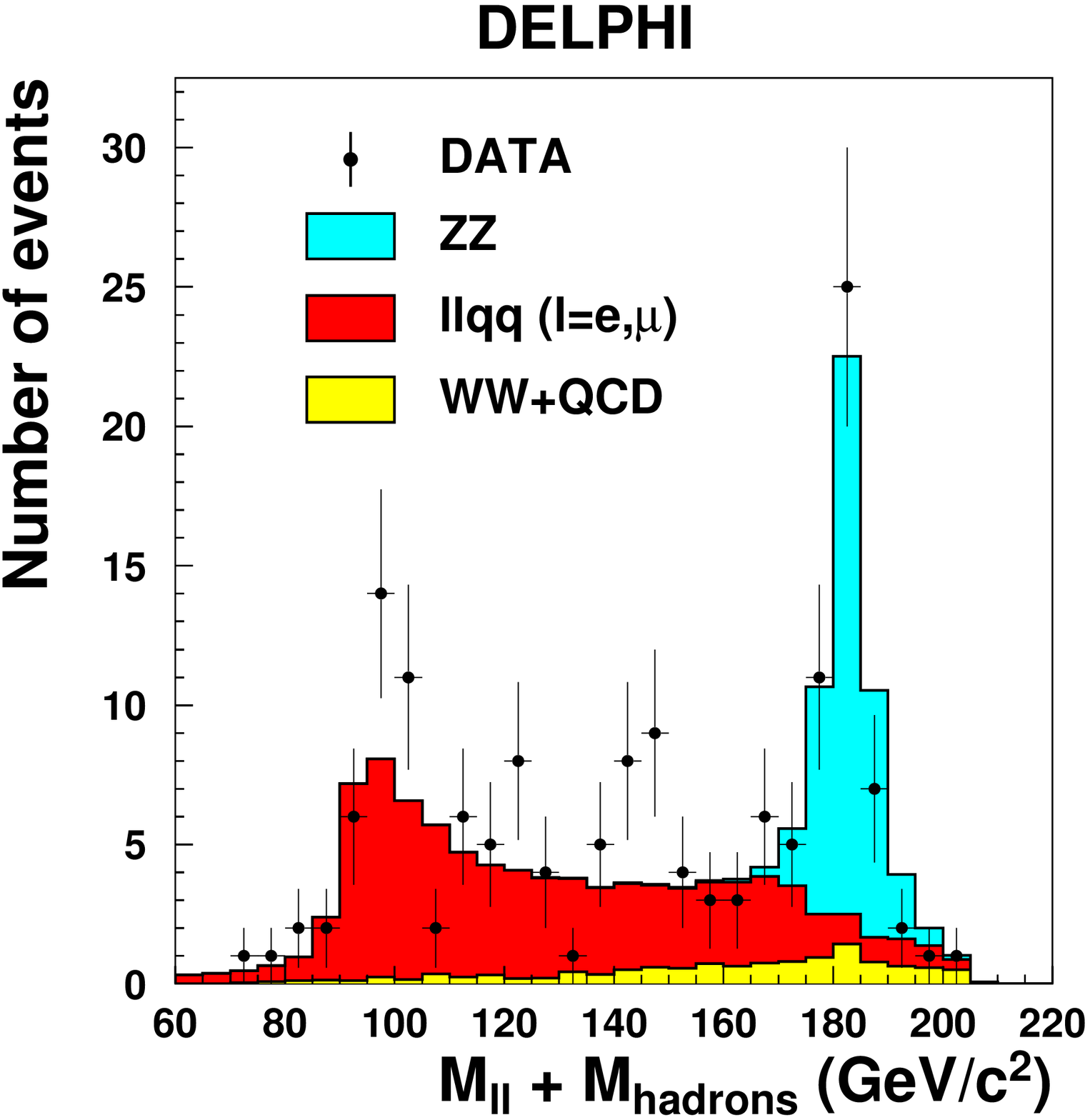}}
    \caption{
      Distribution of the mass of the lepton pair (hadron system) when 
      that of the hadron system (lepton pair) is within 15 GeV/c$^2$
      of the nominal $Z$ mass (upper left and right-hand plots,
      respectively) and of the sum of the masses of the lepton pair 
      and hadron system (lower plot) for the selected events.  
      The dots are the data taken in 
      1997-2000 and the histograms represent the simulation.
      The background labelled $l^+l^-q{\bar q}$ ($l=e,\mu$) corresponds 
      to the predicted contributions from these final states not satisfying
      the signal definition described in Section 3.
      }
    \label{llqq:fig4}
  \end{center}
\end{figure}

The distribution of the mass of one fermion pair ($M_{ll}$
or $M_{q \bar{q}}$) when the mass of the second pair is within 
15 GeV/c$^2$ of the nominal $Z$ mass is shown in Figure \ref{llqq:fig4}
(upper left and right-hand plots, respectively) for the selected events. 
The corresponding distribution of the sum of the masses of the two fermion-pairs is shown in
the lower plot of Figure \ref{llqq:fig4}. The observed distributions are in reasonable
agreement with the predictions from simulation. To select on-shell $ZZ$ 
production, cuts were placed simultaneously on the masses of the 
$l^+ l^-$ pair, on the remaining hadron system and on their sum, 
taking into account in the performance optimization the different 
mass resolutions of these final states and the presence of the 
single-resonant process, $e^+ e^- Z$, in the case of the
$e^+e^- q \bar{q}$ channel. The observed and predicted numbers of selected 
on-shell $ZZ$ events are shown in Table~\ref{table:llqq}.
The residual background is composed of $l^+l^- q \bar{q}$ events 
generated outside the signal region (see Section 3), of other processes, principally 
$W^+W^-$, other $ZZ$ decays and, in the case of $e^+e^- q \bar{q}$,
$q \bar{q} (\gamma)$ production. 
These results were used to derive the combined values of the 
NC02 cross-section described in Section 10.

\begin{table}[htp]
\begin{center}
\begin{tabular}
{|c|c|c|c|c|c|}
\hline
\multicolumn{6}{|c|}{ \ \ \ } \\
\multicolumn{6}{|c|}{ $ZZ \rightarrow \mu^+ \mu^- q \bar {q}$ } \\
\multicolumn{6}{|c|}{ \ \ \ } \\
\hline
$\sqrt s$   & Integrated              & Selection       & Predicted  & Predicted & Selected  \\ 
$[$GeV$]$   & luminosity [pb$^{-1}$]  & efficiency      & background & total MC  & data      \\ 
\hline
183            & 54.0 & 0.88 & 0.04 & 0.52 & 3  \\ 
189            &158.1 & 0.87 & 0.22 & 4.14 & 5  \\ 
192            & 25.8 & 0.89 & 0.05 & 0.91 & 0  \\ 
196            & 76.9 & 0.88 & 0.12 & 2.71 & 1  \\ 
200            & 84.3 & 0.87 & 0.15 & 3.20 & 5  \\ 
202            & 41.1 & 0.84 & 0.08 & 1.62 & 0  \\ 
205            & 83.3 & 0.87 & 0.17 & 3.34 & 2  \\ 
207            &141.8 & 0.84 & 0.30 & 5.70 & 5  \\ 
\hline
Total          &665.3 & 0.86 & 1.13 &22.14 & 21 \\ 
\hline
\hline
\multicolumn{6}{|c|}{ \ \ \ } \\
\multicolumn{6}{|c|}{ $ZZ \rightarrow e^+ e^- q \bar {q}$ } \\
\multicolumn{6}{|c|}{ \ \ \ } \\
\hline
$\sqrt s$   & Integrated              & Selection       & Predicted  & Predicted & Selected  \\ 
$[$GeV$]$   & luminosity [pb$^{-1}$]  & efficiency      & background & total MC  & data      \\ 
\hline
183            & 54.0 & 0.72 & 0.18 & 0.73 & 0  \\ 
189            &158.1 & 0.74 & 0.64 & 4.19 & 3  \\ 
192            & 25.8 & 0.75 & 0.13 & 0.93 & 1  \\ 
196            & 76.9 & 0.74 & 0.36 & 2.73 & 4  \\ 
200            & 84.3 & 0.74 & 0.45 & 3.27 & 2  \\ 
202            & 41.1 & 0.72 & 0.17 & 1.50 & 1  \\ 
205            & 83.3 & 0.74 & 0.34 & 3.30 & 4  \\ 
207            &141.8 & 0.70 & 0.55 & 5.37 & 4  \\ 
\hline
Total          &665.3 & 0.73 & 2.82 &22.02 & 19 \\ 
\hline
\end{tabular}
\end{center}
\caption[.]{
\label{table:llqq}
Integrated luminosities, selection efficiencies and number of 
observed and expected selected events in the $ZZ \rightarrow \mu^+ \mu^- q \bar{q}$ 
and $ZZ \rightarrow e^+ e^- q \bar{q}$ channels.
}
\end{table}

\subsection{Systematic errors}

Several possible sources of systematic errors were checked. Uncertainties in the 
efficiency of the lepton identification were studied comparing semileptonic $WW$ events 
selected in data and simulation using the strong lepton identification 
criteria. Uncertainties in signal efficiencies from the description of the kinematic 
observables used were evaluated comparing the $P_t$ and $\chi^2$ distributions 
in data and simulation for all $l^+l^- q \bar{q}$ events selected without 
mass cuts. 
Corresponding uncertainties in background levels were evaluated by comparing 
samples of events selected in data and in simulation, requiring both isolated 
tracks to be not identified as leptons, while maintaining all the other criteria.
Finally, uncertainties in the background level in the $e^+e^- q \bar{q}$ channel 
from fake electrons were studied with $q{\bar q}(\gamma)$ events selected in data 
and in simulation with purely kinematic criteria. These effects and the statistical
uncertainty of simulated data yielded combined systematic errors on the 
efficiency to select $\mu^+\mu^- q \bar{q}$ and $e^+e^- q \bar{q}$ events of 
$\pm$3.0\%, and on the relative uncertainty in the background level of 
$\pm$15\%\footnote{These evaluations were limited in accuracy by the statistics of the 
available samples.}. 
Finally, additional errors of $\pm$2\% and $\pm$3\% were added in the procedure
to extract and combine the cross-sections with other channels (see Section 10), to 
account for uncertainties from the signal definition used (see Section 3), arising 
partly because of interference effects and because of biases resulting from the 
single-resonant $(Z/\gamma^*) e^+e^-$ process, respectively.

%
%
%
\section {Jets and a pair of isolated $\tau$-leptons \label{sub:QQTT}}
%
%
%

The $\tau^+ \tau^- q \bar{q}$ decay mode 
represents 4.7\% of the $ZZ$ final states. 
The search for $\tau^+ \tau^- q \bar{q}$ final states was based on the
inclusive selection of isolated clusters of particles with low
charged multiplicity and low invariant mass. Each pair
of such clusters was considered as a possible $\tau^+\tau^-$ candidate,
with all other particles assumed to originate from a $q\bar{q}$ system.
Assuming this topology, a set of discriminating variables
reflecting the isolation of the clusters and the likelihood of 
the $\tau^+\tau^-q\bar{q}$ hypothesis was defined. All these variables
were combined into a single quantity using a probabilistic method based on 
likelihood ratio products.
Cutting on this combined variable enabled $\tau^+\tau^-q\bar{q}$
final states consistent with $ZZ$ production to be selected with 
high efficiency and low background contamination.

\subsection{Definition of $\tau$-clusters}
\label{tt-sec12}

Each charged particle with momentum exceeding 1~GeV/c was considered
as a $\tau$-cluster candidate. Every other charged particle (with 
energy exceeding 1~GeV) and photon (with energy exceeding 0.5~GeV) was 
then successively added to each one of these candidates and the 
incremental change in the cluster mass $\Delta M_{cl}$ was computed.
After sorting the particles in increasing order of $\Delta M_{cl}$, the particle 
which produced the minimal $\Delta M_{cl}$ was kept if 
the charged and neutral particle cluster multiplicities were each less than 4,
if $\Delta M_{cl}$ was less than 1.5~GeV/c$^2$
and, for clusters with one (more than one) 
charged particle, if their mass, $M_{cl}$, 
was less than 2.5 (1.9)~GeV/c$^2$.

After adding each particle, the determination of $\Delta M_{cl}$ was repeated 
for all remaining particles with the modified cluster. The procedure was 
continued until all combinations had been tried. If the initial charged 
particle was identified as an electron or a muon (using the standard packages
described in \cite{DELPHIPER}), only photons were added. To study the isolation 
of these $\tau$-cluster candidates, all the particles of the events were 
clustered into jets with the JADE algorithm~\cite{lund} (using $y_{min}=0.01$).
The four-momenta of the tracks of the $\tau$-cluster were subtracted from the 
jets to which they belonged, and the nearest jet with a remaining energy exceeding 
3 (1.5)~GeV, for a cluster with one (more than one) charged particle, 
was determined. Two variables characterising the isolation of each cluster 
were defined: the angle $\theta_{cl}$ between the cluster and the nearest jet, and 
the transverse momentum $P_t$ of the most energetic particle in the cluster 
with respect to the nearest jet.

\subsection{Pre-selection}

Events were required to have at least 7 reconstructed charged particles and a 
charged energy greater than 0.30 $\sqrt{s}$. To suppress radiative returns to 
the $Z$, events were rejected if a photon with energy exceeding 55\% 
of the beam energy was found. At least two $\tau$-clusters per event
were required, satisfying the isolation criterion $\cos\theta_{cl}<0.95$
(0.92), for clusters with one (more than one) charged particle, 
having a sum of charges satisfying the condition $|Q_1+Q_2|<2$, and with at 
least one of these clusters containing only one charged particle.

All particles not used to define the two $\tau$ candidates were then
clustered into jets and a kinematic fit~\cite{kinfit} with six constraints, 
consisting of four-momentum conservation and imposition of the condition
$M_{\tau^+\tau^-}$ = $M_{q\bar{q}}$ = $M_Z$, was applied to the event, where
$M_{\tau^+\tau^-}$ and $M_{q\bar{q}}$ are the invariant masses of the $\tau$
and jet systems, respectively. To exploit 
the large missing energy in the $\tau$ decay, the missing energies of each 
$\tau$-cluster, $D_E^{1,2}$ (defined as the differences 
between each fitted cluster energy, $E^{1,2}_{fit}$, 
and the corresponding measured value), were then required to satisfy:

\begin{itemize}
\item
min($D_E^1,D_E^2) > -15$ GeV,
\item
min($D_E^1,D_E^2) > 10$~GeV if the clusters contained 
identified leptons (electrons or muons) of the same type,
\item
$D_E^i > 3$~GeV ($i=1,2$) for $\tau$-clusters containing an identified lepton, 
\item
$D_E^1+D_E^2 > 10$~GeV if one or both $\tau$-clusters contained 
an identified lepton.
\end{itemize}

After all these pre-selection criteria, the total number of events
observed in data and predicted in simulation were 505 and 460.7, respectively.



\subsection{Probabilistic selection}

\begin{figure}[h]
  \begin{center}
    \mbox{\epsfysize=7.8cm\epsffile{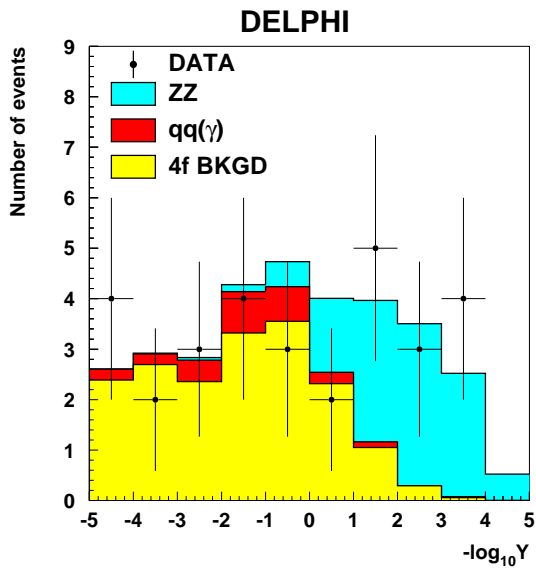}}
    \caption{Distribution of the combined discriminating variable for the
      $\tau^+\tau^-q\bar{q}$ analysis, obtained after 
      enriching with signal events by requiring
      $M_{\tau^+\tau^-}$ and $M_{q\bar{q}}$
      to be compatible with the $Z$ boson mass.
      The dots are the data taken in 1997-2000 and the  
      histograms represent the simulation.
      }
    \label{tt-fig1}
  \end{center}
\end{figure}

A probabilistic method based on likelihood ratio products 
was used to combine several
variables defined at the level of the $\tau$-clusters
($E^{1,2}_{fit}$, $D_E^{1,2}$, $\cos\theta_{cl}^{1,2}$,  
$P_t^{1,2}$), and the $\chi^2$ probabilities of three kinematic fits,
performed using the constraints from four-momentum conservation,
using in addition the hypothesis of $ZZ$ production (as described in 
section 7.2) or that of $WW$ production (requiring the two 
di-jet masses reconstructed in the pairing with smallest $\chi^2$
to be equal to $M_W$). 
The probability density functions were determined for each quantity
by combining the predictions from all centre-of-mass energies separately 
for $\tau$-clusters with leptons, with one charged particle and
with more than one charged particle. The separation power 
of $Y$, the combined discriminant variable obtained, 
is illustrated in Figure~\ref{tt-fig1}, where the 
distribution of $-\log_{10}Y$ is shown for a sample enriched with 
signal events by requiring $M_{\tau^+\tau^-}$ and $M_{q\bar{q}}$ 
to be compatible with the $Z$ boson mass.

\subsection{Results}

\begin{figure}[h]
  \begin{center}
    \mbox{\epsfysize=7.8cm\epsffile{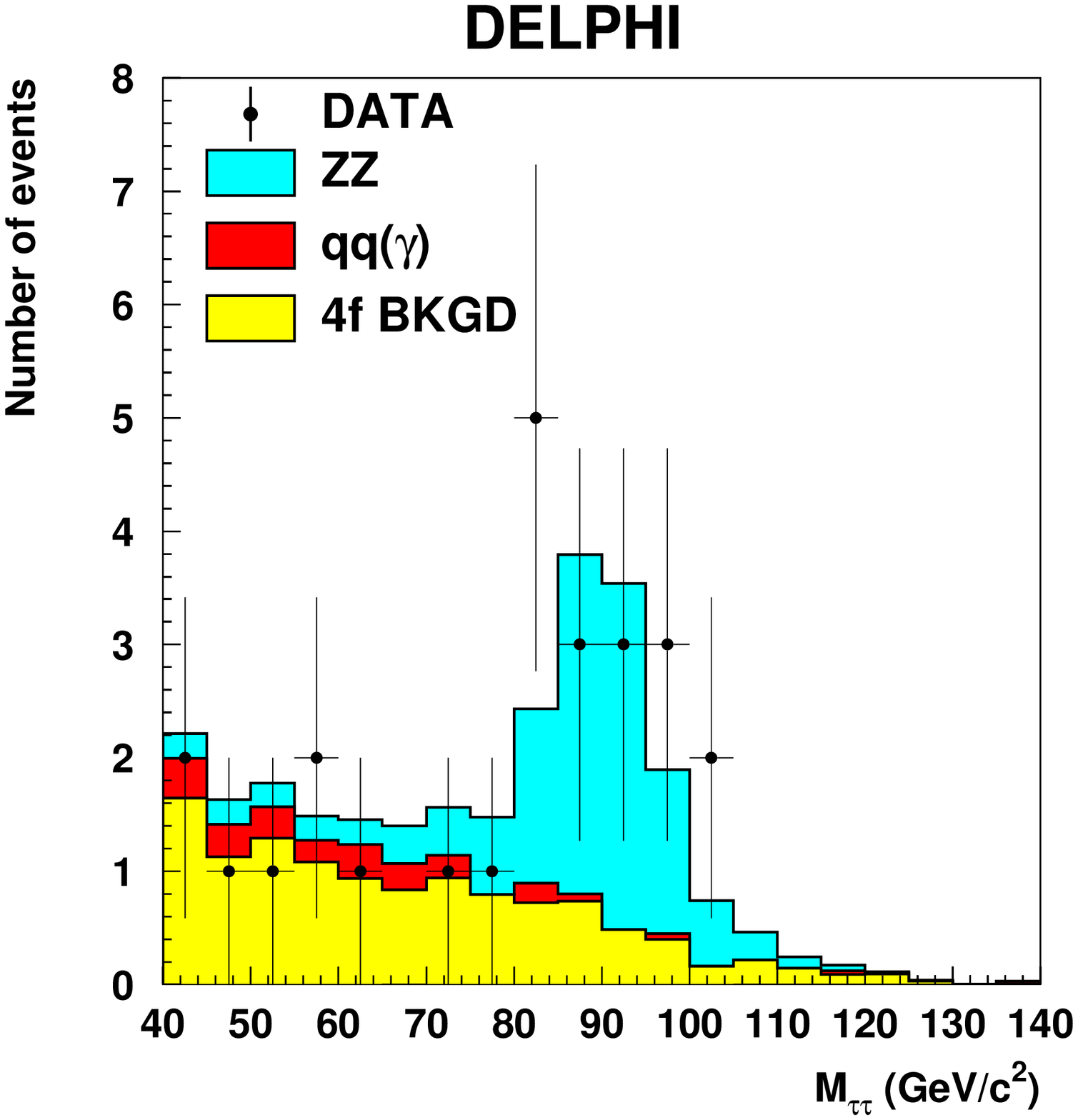}}
    \mbox{\epsfysize=7.8cm\epsffile{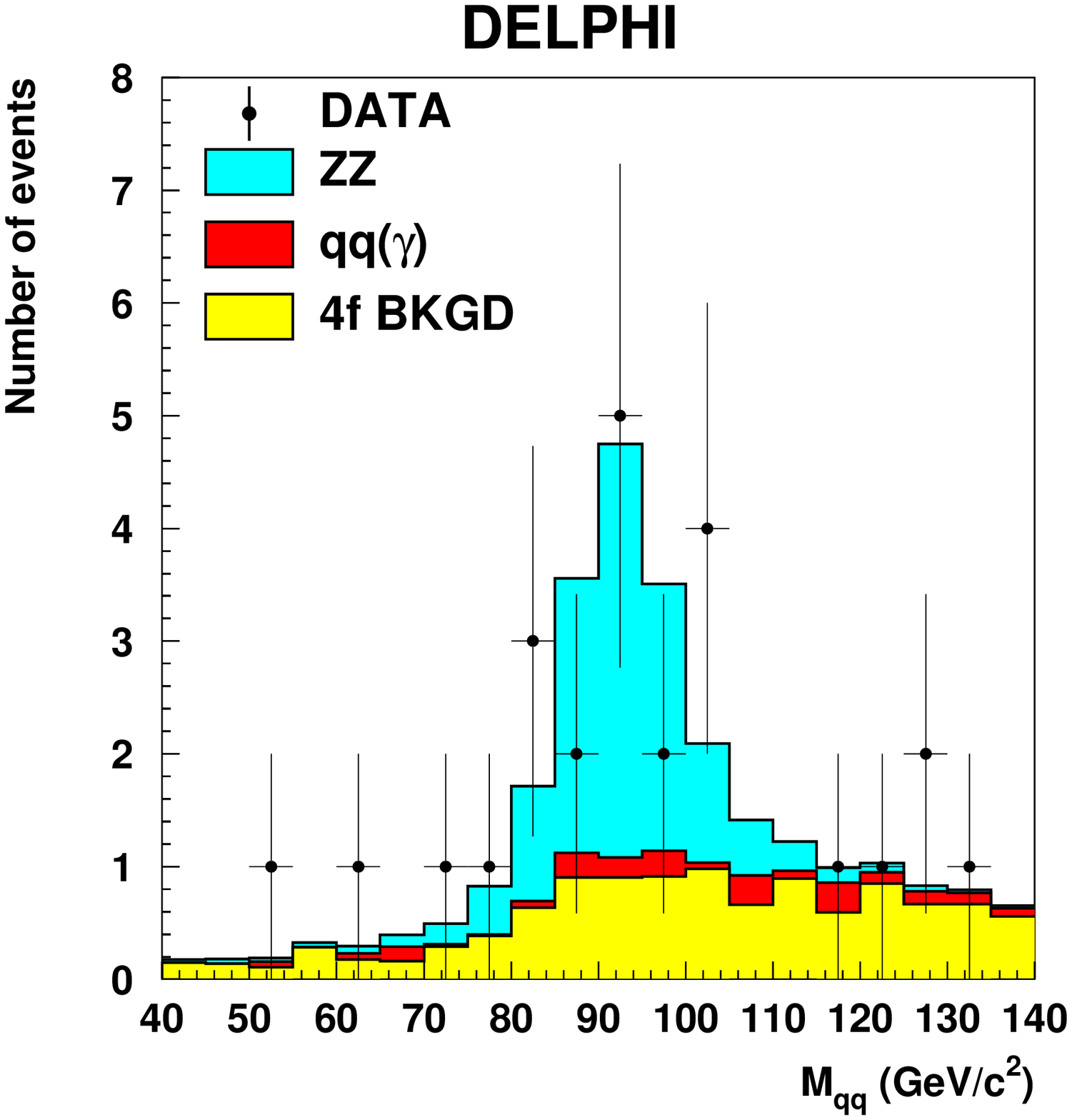}}
    \caption{Distribution of $M_{\tau^+\tau^-}$ and $M_{q\bar{q}}$,
      obtained through kinematic fitting of the event 
      with the constraints from four-momentum conservation only,
      for the selected events satisfying $-\log_{10}Y>0.5$ (see the text).
      The dots are the data taken in 1997-2000 and the  
      histograms represent the simulation.
      }
    \label{tt-fig2}
  \end{center}
\end{figure}

To isolate the on-shell $ZZ$ production process, a cut, 
chosen to maximise the product of the efficiency and purity,
was applied to the combined variable: $-\log_{10}Y>0.5$. 
The fermion pair mass distributions for the events selected by this cut 
($M_{\tau^+\tau^-}$ and $M_{q\bar{q}}$) are shown in Figure~\ref{tt-fig2}, 
obtained after a kinematic fit of the event 
with the constraints from four-momentum conservation only.
The observed distributions are in reasonable
agreement with the predictions from simulation.
Additional mass cuts were used to improve the signal-to-background ratio 
for the final selection:
$75 < M_{\tau^+\tau^-,q\bar{q}} < 110$ GeV/c$^2$ and
$170 < M_{\tau^+\tau^-} + M_{q\bar{q}} < 200$ GeV/c$^2$.
The observed and predicted numbers of selected events are shown in 
Table~\ref{tt-tab3}. 
The background is composed of $q \bar{q}q \bar{q}$, 
$(e^+e^-,\mu^+ \mu^-) q \bar{q}$ and $q \bar{q} (\gamma)$ events. The 
$q \bar{q}q \bar{q}$ final states were primarily from $WW$ 
decays. The overall contribution from other $ZZ$ decays was 
less than 30\% of the total background. 
These results were used to derive the combined values of the 
NC02 cross-section described in Section 10. 

\begin{table}[h]
\begin{center}
\begin{tabular}
{|c|c|c|c|c|c|}
\hline
\multicolumn{6}{|c|}{ \ \ \ } \\
\multicolumn{6}{|c|}{ $ZZ \rightarrow \tau^+ \tau^- q \bar {q}$ } \\
\multicolumn{6}{|c|}{ \ \ \ } \\
\hline
$\sqrt s$   & Integrated              & Selection       & Predicted  & Predicted & Selected  \\ 
$[$GeV$]$   & luminosity [pb$^{-1}$]  & efficiency      & background & total MC  & data      \\ 
\hline
183            & 54.0 & 0.28 & 0.09 & 0.28 & 0  \\ 
189            &158.1 & 0.34 & 0.52 & 2.08 & 1  \\ 
192            & 25.8 & 0.40 & 0.12 & 0.49 & 0  \\ 
196            & 76.9 & 0.41 & 0.32 & 1.56 & 0  \\ 
200            & 84.3 & 0.41 & 0.42 & 1.91 & 1  \\ 
202            & 41.1 & 0.41 & 0.20 & 0.94 & 2  \\ 
205            & 83.3 & 0.41 & 0.38 & 1.91 & 4  \\ 
207            &141.8 & 0.40 & 0.64 & 3.26 & 5  \\ 
\hline
Total          &665.3 & 0.38 & 2.69 &12.43 & 13 \\ 
\hline
\end{tabular}
\end{center}
\caption[.]{
\label{tt-tab3}
Integrated luminosities, selection efficiencies and number of 
observed and expected selected events in the $ZZ \rightarrow \tau^+ \tau^- q \bar{q}$ 
channel, after cuts on $M_{\tau^+\tau^-}$, $M_{q\bar{q}}$ and on their sum. 
}
\end{table}

\subsection{Systematic errors}

The main sources of systematic error in the selection of
$ZZ \rightarrow \tau^+ \tau^- q \bar{q}$ events arose from 
uncertainties in the reconstruction of the $\tau$-clusters
and associated isolation variables. 
After the pre-selection, 
about 10\% more events were selected in data as compared to 
Monte Carlo (see Section 7.2). At this level the selection
is dominated by $WW$ processes where one $W$ decays into $\tau \nu_{\tau}$ and 
the other into hadrons. Although the corresponding topology is not identical with 
that of the signal, it is similar since it contains a $\tau$-lepton to be identified 
in the presence of hadronic jets. A systematic uncertainty of $\pm$10\% on the selection 
efficiency was assigned to take this discrepancy into account and 
based on studies of uncertainties in the lepton identification used in the 
selection of the $\tau$-clusters.
An uncertainty in the background level of $\pm$30\%
was also specified because the fraction of background events originating from other 
$ZZ$ decays (see Section 7.4) was neglected in the procedure to derive combined NC02 
cross-sections. Both errors were assumed to be fully correlated between 
the centre-of-mass energies analysed.

\section{Final states with four charged leptons\label{sub:LLLL}}
%
%
%

The $l^+l^-l^+l^-$ decay modes with $l=e,\mu$ comprise 0.4\% of the 
$ZZ$ final states. The event topology is clean and the only significant 
background comes from non-resonant $e^+e^-l^+l^-$ production. 

\subsection{Selection procedure}

Events were first selected if they contained between 4 and 8 charged particles, 
accompanied by at most 10 neutral particles, irrespective of particle identification.
In order to take into account bremsstrahlung effects for candidate electrons, the 
momentum of a charged particle was replaced by the sum of energies of electromagnetic 
clusters in a narrow cone around the track direction if it gave a larger value. 
The total invariant mass of the charged particles had to be greater than 
50 GeV/c$^2$, and the minimum invariant mass after discarding any one of the charged 
particles larger than 20 GeV/c$^2$. All combinations of four charged particles with 
total charge zero were then examined, and a combination was selected if all the
following criteria were satisfied:

\begin{itemize}
\item
all four tracks had impact parameters at the interaction point 
smaller than 3.0 and 0.5 cm, respectively in the projections containing 
the beam axis and perpendicular to it, and polar angles between 
$10^\circ$ and $170^\circ$,
\item
at least three of the four charged particles had momenta greater than 
5 GeV/c, and the least energetic particle had momentum greater than 2 GeV/c, 
\item
a system of two oppositely charged particles was found with both their
invariant and missing masses within 10 GeV/c$^2$ of the $Z$ boson mass 
and having the same flavour (when both were identified as either 
electrons or muons using the standard packages described in \cite{DELPHIPER}), 
\item
the two particles complementary to this system were separated by at 
least $90^\circ$ from each other, 
\item
the invariant mass of all pairs of oppositely charged particles in the event 
exceeded 5~GeV/c$^2$.
\end{itemize}

\subsection{Results}

\begin{figure}[t]
\begin{center}
\mbox{\epsfysize=11.0cm\epsffile{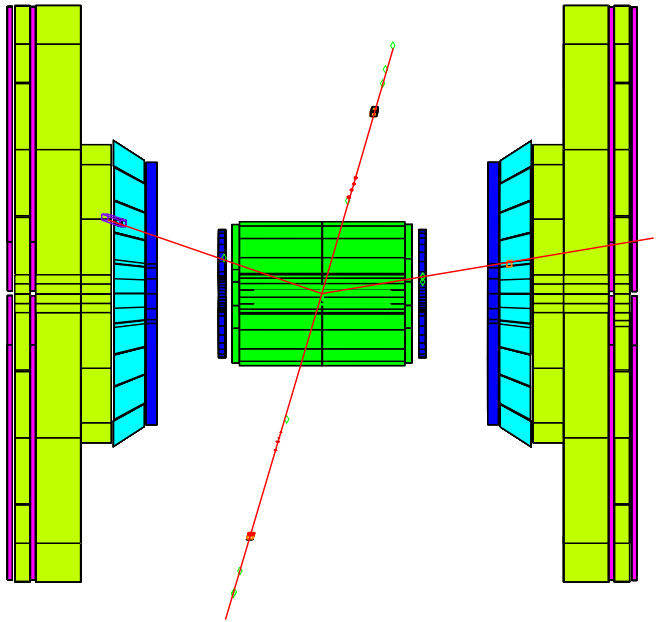}}
\caption{
The $ZZ \to \mu^+ \mu^- \mu^+ \mu^-$ candidate selected at 199.5~GeV 
centre-of-mass energy. Three muons are observed in the barrel 
and one in the forward part of the detector (the muon track going 
to the left is out of the plane). 
A kinematic fit was performed assuming the production of two identical 
heavy objects, each decaying into $\mu^+ \mu^-$. Out of
two possible combinations to form the $\mu^+ \mu^-$ pairs 
the highest probability was found for the top-bottom and left-right pairs.
The corresponding fitted invariant mass, 90.8~GeV/c$^2$, 
is fully compatible with $ZZ$ production.  
}
\label{fig:mmmm}
\end{center}
\end{figure}

The observed and predicted numbers of selected events are shown in 
Table~\ref{rllll}. They are in overall agreement within the statistical errors. 
These results were used to derive the combined values of the NC02 
cross-section described in Section 10. The two events selected at 
$\sqrt{s}$ = 188.6 and 199.5 GeV were in the $\mu^+\mu^-e^+e^-$
and $\mu^+\mu^-\mu^+\mu^-$ final states, respectively. An event 
display for the latter is shown in Figure~\ref{fig:mmmm}.

\vskip0.2cm

\begin{table}[htp]
\begin{center}
\begin{tabular}
{|c|c|c|c|c|c|}
\hline
\multicolumn{6}{|c|}{ \ \ \ } \\
\multicolumn{6}{|c|}{ $ZZ \rightarrow l^+l^-l^+l^-$ } \\
\multicolumn{6}{|c|}{ \ \ \ } \\
\hline
$\sqrt s$   & Integrated              & Selection       & Predicted  & Predicted & Selected  \\ 
$[$GeV$]$   & luminosity [pb$^{-1}$]  & efficiency      & background & total MC  & data      \\ 
\hline
183            & 54.0 & 0.53 & 0.09 & 0.17 & 0  \\ 
189            &158.1 & 0.62 & 0.15 & 0.76 & 1  \\ 
192            & 25.8 & 0.58 & 0.06 & 0.19 & 0  \\ 
196            & 76.9 & 0.54 & 0.13 & 0.47 & 0  \\ 
200            & 84.3 & 0.54 & 0.17 & 0.60 & 1  \\ 
202            & 41.1 & 0.54 & 0.07 & 0.27 & 0  \\ 
205            & 83.3 & 0.50 & 0.22 & 0.62 & 0  \\ 
207            &141.8 & 0.41 & 0.28 & 0.84 & 0  \\ 
\hline
Total          &665.3 & 0.53 & 1.17 & 3.92 & 2  \\ 
\hline
\end{tabular}
\end{center}
\caption[.]{
\label{rllll}
Integrated luminosities, selection efficiencies and number of 
observed and expected selected events in the $ZZ \rightarrow l^+l^-l^+l^-$ 
channel.
}
\end{table}

%
%

\subsection{Systematic errors}

The main sources of systematic error in the analysis of the $ZZ \rightarrow l^+l^-l^+l^-$ 
decay modes with $l=e,\mu$ were from the statistics of the Monte Carlo samples produced to 
compute the residual background and the selection efficiency and from approximations used 
to translate the results in terms of NC02 cross-sections (see Section 3). 
These three uncertainties were propagated into the combination described in Section 10. 
They amounted to $\pm$12\%, $\pm$5\% and $\pm$5\% relative errors on the cross-sections 
obtained in this channel, respectively, and dominated over uncertainties in 
background cross-sections and reconstruction. The third error was assumed to be fully 
correlated between the energies analysed. 

%
%
%
\section{Two isolated electrons or muons with missing energy}
%
%
%

The $\nu \bar{\nu} l^+ l^-$ decay modes with $l=e,\mu$ comprise 2.7\% of the $ZZ$ final 
states. The events are characterized by two acollinear charged leptons 
of the same flavour, with both invariant and recoil masses close to the
$Z$ mass, and by large missing energy. Although it has different production
kinematics, the $WW$ process also contributes to these final states with a 
large cross-section. A significant fraction of the corresponding events have 
exactly the same features and constitute a substantial background.

\subsection{Selection procedure}

To ensure good reconstruction, tracks were required to have impact parameters 
at the interaction point smaller than 3.0 and 0.5 cm, respectively in the 
projections containing the beam axis and perpendicular to it, and polar angles 
between $20^\circ$ and $160^\circ$.
Events with two particles identified as $e^+e^-$ or $\mu^+\mu^-$ (using the looser 
identification criteria described in \cite{DELPHIPER} and calorimetric
requirements), were selected if their total energy was less than 60\% of $\sqrt s$, 
if the angle between them was in the range [$\theta_{min}$,$170^\circ$],
where $\theta_{min} = 0.95 \arccos{(1 - 8m_Z^2/s)}$, if the polar angle of 
the event missing momentum was between $25^\circ$ and $155^\circ$,
and if the reconstructed invariant masses satisfied:

\begin{center}
min\{$|~M_{Z} - m(l^+l^-)~|~$,$~|~M_{Z} - m_{recoil}(l^+l^-)~|$\} 
      $<$ 4 GeV/c$^2$, 
max\{$|~M_{Z} - m(l^+l^-)~|~$,$~|~M_{Z} - m_{recoil}(l^+l^-)~|$\} 
      $<$ 8 GeV/c$^2$,
\end{center}

\noindent 
where $m_{recoil}(l^+l^-)$ is the invariant mass recoiling 
against the $l^+l^-$ pair. Final state radiation for candidate electrons 
was taken into account as discussed in Section 8.1 for the four-lepton analysis. 

\subsection{Results}


The observed and predicted numbers of selected events are shown in 
Table~\ref{rnnll}. They are in overall agreement within the statistical errors. 
These results were used to derive the combined 
values of the NC02 cross-section described in Section 10. 

\begin{table}[htp]
\begin{center}
\begin{tabular}
{|c|c|c|c|c|c|}
\hline
\multicolumn{6}{|c|}{ \ \ \ } \\
\multicolumn{6}{|c|}{ $ ZZ \rightarrow \nu \bar{\nu} \mu^+ \mu^- , 
\nu \bar{\nu} e^+ e^-$ } \\
\multicolumn{6}{|c|}{ \ \ \ } \\
\hline
$\sqrt s$   & Integrated              & Selection       & Predicted  & Predicted & Selected  \\ 
$[$GeV$]$   & luminosity [pb$^{-1}$]  & efficiency      & background & total MC  & data      \\ 
\hline
183            & 54.0 & 0.30 & 0.31 & 0.42 & 0  \\ 
189            &158.1 & 0.32 & 0.80 & 1.65 & 2  \\ 
192            & 25.8 & 0.35 & 0.16 & 0.34 & 2  \\ 
196            & 76.9 & 0.33 & 0.74 & 1.32 & 2  \\ 
200            & 84.3 & 0.32 & 0.72 & 1.39 & 1  \\ 
202            & 41.1 & 0.29 & 0.36 & 0.68 & 0  \\ 
205            & 83.3 & 0.27 & 0.58 & 1.18 & 1  \\ 
207            &141.8 & 0.25 & 1.00 & 1.94 & 2  \\ 
\hline
Total          &665.3 & 0.30 & 4.67 & 8.92 & 10 \\ 
\hline
\end{tabular}
\end{center}
\caption[.]{
\label{rnnll}
Integrated luminosities, selection efficiencies and number of 
observed and expected selected events in the $ZZ \rightarrow \nu\bar{\nu}\mu^+\mu^-,
\nu\bar{\nu}e^+ e^-$ channels.
}
\end{table}

\subsection{Systematic errors}

The main sources of systematic error in the analysis of the $ZZ \rightarrow \nu \bar{\nu} 
l^+ l^-$ decay modes with $l=e,\mu$ were from the statistics of the Monte Carlo samples 
produced to compute the remaining background and the selection efficiency and from 
uncertainties in the background level. These three uncertainties were propagated into the 
combination described in Section 10.  They amounted to $\pm$13\%, $\pm$5\% and $\pm$5\%
relative errors on the cross-sections obtained in this channel, respectively, and dominated 
over experimental errors in the selection efficiency and over uncertainties from  
approximations used to translate the results in terms of NC02 cross-sections. 
The third error was assumed to be fully correlated between the energies analysed.

%
%
%
\section{Combined NC02 cross-sections}
%
%
%

The measurements described in the previous sections for the different $ZZ$ 
channels show reasonable agreement between the data and simulation. 
In this section the comparison is presented in terms of extracted 
NC02 cross-sections combined over all channels at each centre-of-mass energy, 
as well as over all energies by normalising to the SM predictions, both separately 
in each channel, and for all channels together. Such combinations enable more 
comprehensive and statistically meaningful checks of the SM prediction.
Cross-sections for individual sub-channels are also given at each energy in 
the case of the most important $ZZ$ final states. 

\subsection{Likelihood method}

To obtain the cross-section information in the different cases, probability 
functions at each energy and for each sub-channel, defined with respect to variations 
of the value of the NC02 cross-section, were combined into global likelihoods. 
For the channels $q \bar{q} q \bar{q}$ and $\nu \bar{\nu} q\bar{q}$ (analysed in 
the IDA stream), the shapes of the probability functions were derived from the fits 
to the distributions of the combined variables defined in each  
analysis. For the remaining channels the Poissonian probabilities were constructed, 
based on the numbers of events selected in the data and predicted in the simulation. 
The central value for each measurement was defined as the median of 
the corresponding likelihood. The statistical errors were obtained from the two intervals 
on either side of the median each containing 34.13\% of the total probability. In cases 
when less than 31.74\% of the probability was below the maximum of the likelihood, an upper 
limit was given, defined as the value which was exceeded by 5\% of the integrated probability
distribution.

\subsection{Propagation of systematic errors}

The propagation of systematic uncertainties affecting each different final state to 
its probability function was studied by introducing appropriately modified assumptions 
for backgrounds and efficiencies (as described in the corresponding sections). 
The following procedure was applied to propagate all systematic errors in the combination 
of any selection of channels or energy points. Uncertainties were divided into four groups 
according to their correlation among energies and channels.
The only significant channel-correlated uncertainties were on the integrated luminosities 
collected at each centre-of-mass energy. They amounted to $\pm$0.3\% (theory error) and 
$\pm$0.6\% (measurement error) 
and were respectively treated as fully correlated and 
uncorrelated among the energies. 

A set of Gaussian random numbers was then drawn and used 
to assign values for each kind of error taking into account its nature (common random number 
among channels or energies in cases of fully correlated errors, and independent ones 
otherwise). The probability functions were then modified accordingly and their combination 
was repeated a large number of times. The standard deviation of the Gaussian-like 
distribution of central values obtained was taken to represent the total systematic error 
affecting the combined measurement for the selection of channels or energy points considered.

\begin{figure}[h]
  \begin{center}
    \mbox{\epsfxsize=9.8cm\epsfysize=9.8cm\epsffile{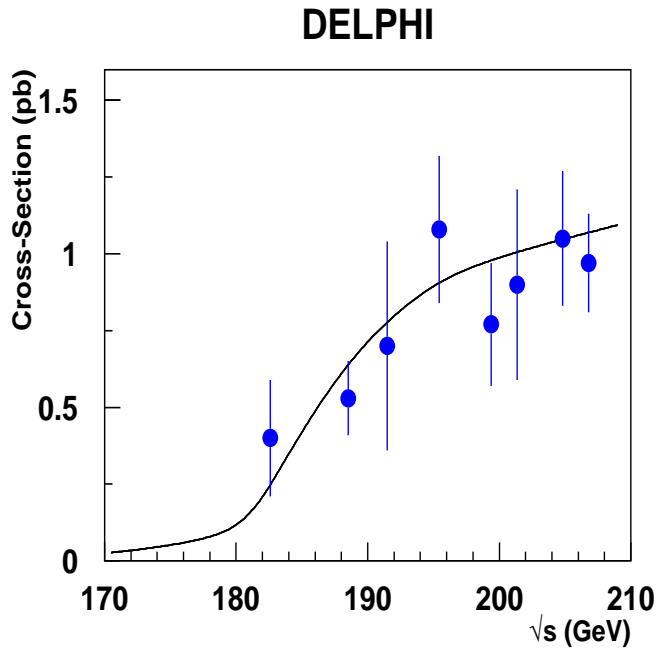}}    
  \caption{ Combined NC02 cross-sections measured from data collected in
            1997-2000. The errors shown are sums in quadrature of the statistical
            and systematic errors. The solid curve is the SM prediction computed using 
            the {\tt YFSZZ} generator~\cite{lancon}. The corresponding uncertainty 
            is about $\pm 2$\%~\cite{ws.gene}. Values obtained with the {\tt EXCALIBUR} 
            generator were consistent with these predictions within 1\%.
           \label{fig:zzcro183_208}}
  \end{center}
\end{figure}

\begin{table}
\begin{center}
\begin{tabular}{|c|c|c|c|} 
\hline
$\sqrt s$   & $\sigma_{\tiny NC02}$ &  SM prediction           &  Expected weight \\ 
$[$GeV$]$   & [pb]                  &  [pb]                    &  [\%]            \\
\hline 
 183 & $0.40^{+0.21}_{-0.16} \pm 0.02$ & 0.25                  &  1.8   \\
 189 & $0.53^{+0.12}_{-0.11} \pm 0.02$ & 0.65                  & 17.4   \\
 192 & $0.70^{+0.37}_{-0.31} \pm 0.02$ & 0.78                  &  3.6   \\
 196 & $1.08^{+0.25}_{-0.22} \pm 0.02$ & 0.90                  & 12.7   \\
 200 & $0.77^{+0.21}_{-0.18} \pm 0.02$ & 0.99                  & 15.1   \\
 202 & $0.90^{+0.33}_{-0.29} \pm 0.02$ & 1.00                  &  7.3   \\
 205 & $1.05^{+0.23}_{-0.20} \pm 0.02$ & 1.05                  & 14.8   \\
 207 & $0.97^{+0.16}_{-0.15} \pm 0.02$ & 1.07                  & 27.3   \\
\hline 
\end{tabular}
\end{center}
\caption{ 
\label{tab:xs_ecms}
Measured and predicted NC02 cross-sections at the different energy points.
In the column with measured values, the first errors are statistical and the second 
systematic. The SM prediction was computed using the {\tt YFSZZ} generator~\cite{lancon}.
The uncertainty on these predictions is about $\pm 2$\%~\cite{ws.gene}. Values obtained 
with the {\tt EXCALIBUR} generator were consistent with these predictions within 1\%.
The last column contains the expected weight of each energy point in the global
combination performed over all energies described in the text.
}
\end{table}

\subsection{Results and discussion}

The measured values of the NC02 cross-section at each centre-of-mass energy 
are listed in Table~\ref{tab:xs_ecms} and shown in Figure~\ref{fig:zzcro183_208}, 
together with the SM predictions from the YFSZZ generator~\cite{lancon}.
These measurements, normalised to the predicted SM values, were also combined 
into a single value:

\begin{center}
${\rm R_{\tiny NC02}}$ = 0.91 $ \pm $ $0.08 ~(stat)$ $ \pm $ $0.02 ~(syst)$. 
\end{center}

\noindent  
The systematic errors at individual energies and in the 
combination had the same magnitude because the main uncertainties were fully correlated. 
The measurements at $\sqrt s$ = 183 and 189 GeV were compatible with those previously 
obtained~\cite{delphizz189}. The uncertainty in the SM prediction, of about 
$\pm 2$\%~\cite{ws.gene}, is not shown (as for the other results in this section). 
A good overall agreement with the SM prediction can be observed. The expected weights of 
each measurement in the combination are given in Table~\ref{tab:xs_ecms}. 

\begin{figure}[tbh]
  \begin{center}
    \mbox{\epsfxsize=10.0cm\epsfysize=10.0cm\epsffile{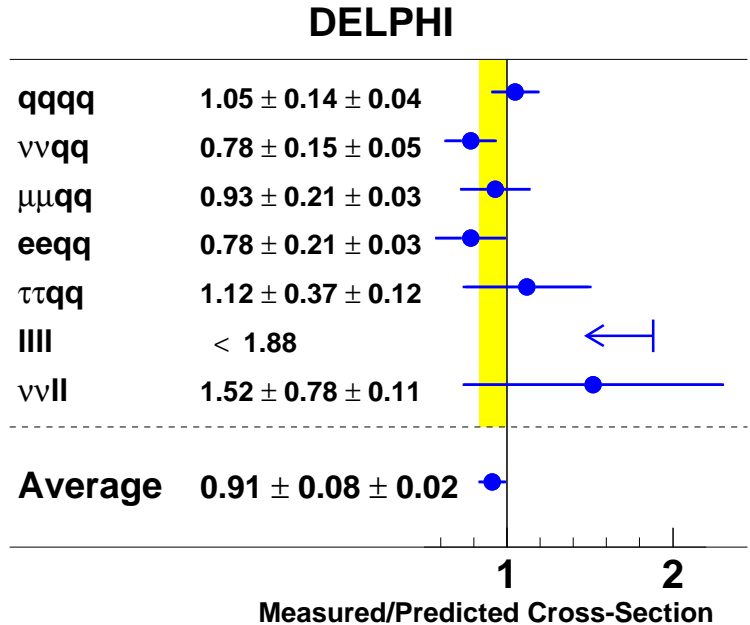}}    
  \caption{ Ratios of measured to predicted cross-sections for
            individual channels combined over all energy points
            (for the two last channels $l=e,\mu$). The predicted 
            cross-sections used are listed in Table~\ref{tab:xs_ecms}. 
            The errors shown are sums in quadrature of the statistical
            and systematic errors. The vertical band displays the total 
            error on the combination of the seven channels. The upper limit 
            given for the $l^+l^-l^+l^-$ channel corresponds to a 95\% confidence level. 
           \label{fig:channel}}
  \end{center}
\end{figure}

The combination of the results from the different energies was also performed for 
each channel separately. The results are listed in Table~\ref{tab:xs_channel}, 
with their expected weight in the global combination, and with the 
corresponding $ZZ$ branching fractions predicted in the SM. The results are also shown 
graphically in Figure~\ref{fig:channel}. The channels were ordered according to their 
expected weight. This order was not always correlated to the decay branching ratios 
because of the different background conditions. As can be seen, the three main channels 
$ZZ \rightarrow q \bar{q} q \bar{q}, \nu \bar{\nu} q \bar{q}$ and $l^+ l^- q \bar{q}$ 
(with $l=e,\mu$) had similar weights, and accounted for over 90\% of the total sensitivity. 
A good overall agreement with the SM predictions can be observed. The combination was also 
performed for the probabilistic $\nu \bar{\nu} q\bar{q}$ cross-check analysis, and the 
result obtained: ${\rm R_{\tiny NC02}} = 0.80^{+0.19}_{-0.18}$ (where the errors are 
statistical), was consistent with that obtained from the main IDA analysis. This channel 
had the largest systematic uncertainty of all (relative to statistical errors). However, 
in this channel as in all others, systematic uncertainties were much smaller than the 
statistical errors. In the four-lepton channel, only an upper limit (quoted at 95\% CL) 
is given. The $\chi^2$ of the seven measurements had a probability of 70\%. 

\begin{table}
\begin{center}
\begin{tabular}{|c|c|c|c|} 
\hline
Channel              &  ${\rm R_{\tiny NC02}}$         & Expected weight  &  Branching Fraction   \\ 
                     &                                 &  [\%]            &  [\%]                 \\
\hline 
$q \bar{q} q \bar{q}$            & $1.05^{+0.14}_{-0.14} \pm 0.04$ &  37.3   &   48.9    \\
$\nu \bar{\nu} q \bar{q}$        & $0.78^{+0.15}_{-0.15} \pm 0.05$ &  24.2   &   28.0    \\
$\mu^+ \mu^- q \bar{q}$          & $0.93^{+0.22}_{-0.20} \pm 0.03$ &  15.6   &    4.6    \\
$e^+ e^- q \bar{q}$              & $0.78^{+0.22}_{-0.19} \pm 0.03$ &  14.3   &    4.6    \\
$\tau^+ \tau^- q \bar{q}$        & $1.12^{+0.40}_{-0.34} \pm 0.12$ &   5.5   &    4.6    \\
$l^+l^-l^+l^-$                   & $ <1.88 $                       &   1.6   &    0.4    \\
$\nu \bar{\nu}l^+l^-$            & $1.52^{+0.86}_{-0.70} \pm 0.11$ &   1.5   &    2.7    \\
\hline 
\end{tabular}
\end{center}
\caption{
\label{tab:xs_channel} 
Ratios of measured to predicted cross-sections for individual channels combined over all 
energy points (for the two last channels $l=e,\mu$). The predicted cross-sections used 
are listed in Table~\ref{tab:xs_ecms}. The upper limit given for the $l^+l^-l^+l^-$ 
channel corresponds to a 95\% confidence level. In the column with measured values, the 
first errors are statistical and the second systematic. The two last columns contain 
the weight of the channel in the global combination over all channels described in the 
text, and the corresponding $ZZ$ SM branching fraction.
}
\end{table}

Finally, for completeness, the values of the measured cross-sections
normalised to the SM predictions are also presented at each energy
in Table~\ref{tab:xs_all} for the three dominant channels. 

\begin{table}
\begin{center}
\begin{tabular}{|c|c|c|c|} 
\hline
$\sqrt s$ & $q \bar{q} q \bar{q}$ & $\nu \bar{\nu} q \bar{q}$ & $l^+ l^- q \bar{q}$    \\ 
$[$GeV$]$ &                       &                           & ($l=e,\mu$)            \\
\hline 
 183      & $<4.21$               & $<3.72$                   & $3.49^{+2.15}_{-1.54}$ \\
 189      & $1.03^{+0.39}_{-0.33}$& $0.59^{+0.34}_{-0.28}$    & $0.98^{+0.41}_{-0.32}$ \\
 192      & $1.34^{+0.84}_{-0.67}$& $<2.27$                   & $<2.83$                \\
 196      & $1.71^{+0.52}_{-0.47}$& $1.36^{+0.54}_{-0.45}$    & $0.93^{+0.51}_{-0.37}$ \\
 200      & $0.78^{+0.35}_{-0.28}$& $0.63^{+0.41}_{-0.34}$    & $1.10^{+0.48}_{-0.37}$ \\
 202      & $1.12^{+0.55}_{-0.45}$& $1.31^{+0.82}_{-0.64}$    & $<1.41$                \\
 205      & $1.17^{+0.37}_{-0.32}$& $0.68^{+0.42}_{-0.34}$    & $0.90^{+0.42}_{-0.33}$ \\
 207      & $1.02^{+0.25}_{-0.22}$& $0.91^{+0.34}_{-0.29}$    & $0.78^{+0.31}_{-0.24}$ \\
\hline 
\end{tabular}
\end{center}
\caption{
\label{tab:xs_all} 
Measured cross-sections normalised to the SM predictions for individual channels at 
individual centre-of-mass energies. The predicted cross-sections used 
are listed in Table~\ref{tab:xs_ecms}. The quoted errors are statistical.
The upper limits shown correspond to a 95\% confidence level.
}
\end{table}

\subsection*{Acknowledgements}
\vskip 3 mm
 We are greatly indebted to our technical 
collaborators, to the members of the CERN-SL Division for the excellent 
performance of the LEP collider, and to the funding agencies for their

support in building and operating the DELPHI detector.\\
We acknowledge in particular the support of \\
Austrian Federal Ministry of Education, Science and Culture,
GZ 616.364/2-III/2a/98, \\
FNRS--FWO, Flanders Institute to encourage scientific and technological 
research in the industry (IWT), Federal Office for Scientific, Technical
and Cultural affairs (OSTC), Belgium,  \\
FINEP, CNPq, CAPES, FUJB and FAPERJ, Brazil, \\
Czech Ministry of Industry and Trade, GA CR 202/99/1362,\\
Commission of the European Communities (DG XII), \\
Direction des Sciences de la Mati$\grave{\mbox{\rm e}}$re, CEA, France, \\
Bundesministerium f$\ddot{\mbox{\rm u}}$r Bildung, Wissenschaft, Forschung 
und Technologie, Germany,\\
General Secretariat for Research and Technology, Greece, \\
National Science Foundation (NWO) and Foundation for Research on Matter (FOM),
The Netherlands, \\
Norwegian Research Council,  \\
State Committee for Scientific Research, Poland, SPUB-M/CERN/PO3/DZ296/2000,
SPUB-M/CERN/PO3/DZ297/2000 and 2P03B 104 19 and 2P03B 69 23(2002-2004)\\
JNICT--Junta Nacional de Investiga\c{c}\~{a}o Cient\'{\i}fica 
e Tecnol$\acute{\mbox{\rm o}}$gica, Portugal, \\
Vedecka grantova agentura MS SR, Slovakia, Nr. 95/5195/134, \\
Ministry of Science and Technology of the Republic of Slovenia, \\
CICYT, Spain, AEN99-0950 and AEN99-0761,  \\
The Swedish Natural Science Research Council,      \\
Particle Physics and Astronomy Research Council, UK, \\
Department of Energy, USA, DE-FG02-01ER41155, \\
EEC RTN contract HPRN-CT-00292-2002. \\

\end{document}